\documentclass[11pt]{article}

\usepackage[english]{babel}

\usepackage{geometry}
\usepackage{footmisc}
\usepackage{appendix}
\usepackage{amssymb}
\usepackage{amsmath}
\usepackage{multirow}
\usepackage{appendix}
\usepackage{float}
\usepackage{graphbox}
\usepackage{authblk}
\usepackage[colorlinks=true, allcolors=blue]{hyperref}
\usepackage{subcaption}
\usepackage{indentfirst}
\usepackage{graphicx}
\usepackage{comment}
\usepackage{multirow}
\usepackage{xcolor}
\usepackage{caption}
\usepackage{cite}
\usepackage{amsfonts}
\usepackage{algorithm}
\usepackage{algorithmic}
\usepackage{rotating} % for sidewaystable / sidewaystable*
\usepackage{booktabs}
\usepackage{array}
\usepackage{xltabular}
\usepackage{booktabs}
\usepackage{longtable}
\usepackage{ragged2e}
\newcolumntype{L}[1]{>{\RaggedRight\arraybackslash}p{#1}}
\usepackage{microtype}
\title{Machine Learning Approaches to Decoding \\ Topological Quantum Codes}
\author[1]{\normalsize Changwon Lee\thanks{\texttt{changwonlee@yonsei.ac.kr}}}
\author[1]{\normalsize Tak Hur\thanks{\texttt{takh0404@yonsei.ac.kr}}}
\author[2]{\normalsize Jeongwoo Jae\thanks{\texttt{jeongwoo.jae@samsung.com}}}
\author[1,3,4]{\normalsize Daniel K. Park\thanks{\texttt{dkd.park@yonsei.ac.kr} (corresponding author)}}
\affil[1]{\small \textit{Department of Statistics and Data Science, Yonsei University, Seoul 03722, Republic of Korea}}
\affil[2]{\small \textit{R\&D center, Samsung SDS, Seoul 05510, Republic of Korea}}
\affil[3]{\small \textit{Department of Applied Statistics, Yonsei University, Seoul 03722, Republic of Korea}}
\affil[4]{\small \textit{Department of Quantum Information, Yonsei University, Seoul 03722, Republic of Korea}}

\date{}

\begin{document}

\maketitle

\vspace{-10mm}
\abstract{Decoding is an essential component of quantum error correction (QEC), translating stabilizer measurement outcomes into corrective actions that suppress logical errors and preserve logical quantum information. Building fault-tolerant architectures requires increasing the code distance, which in turn places growing demands on decoding accuracy, scalability, and practical deployability. While a wide range of decoding algorithms have been proposed and demonstrated, achieving reliable, scalable, and real-time decoding remains a significant challenge. Machine-learning (ML) approaches are particularly well suited to this setting, as quantum error decoding is fundamentally a problem of processing large volumes of classical data with complex spatiotemporal correlations. This chapter surveys ML-based methods for quantum error decoding, with a focus on topological codes and an emphasis on architectural principles, practical performance, and real-time considerations. We first frame decoding as a learning problem and outline key paradigms, including discriminative, generative, and reinforcement-learning formulations. We then introduce the neural network building blocks that underpin most contemporary neural decoders and discuss how these components can be integrated to balance expressivity, scalability, and latency. Building on this architectural perspective, we review recent progress and benchmarks in neural decoding for memory experiments, and discuss real-time decoding, open challenges, and future directions toward scalable fault-tolerant quantum computing.
}

\vspace{0.5em}

\section{Introduction}
\label{sec:1}
% Quantum computing promises computational advantages for problems that are believed to be beyond the practical reach of classical computation, ranging from quantum simulation and cryptography to optimization and scientific computing. Realizing this promise at scale, however, requires quantum processors that can operate reliably despite noise and imperfections. 
% %Physical qubits are affected by imperfect control, environmental decoherence, measurement errors, cross-talk, leakage, and other hardware-specific imperfections. 
% Without active protection and correction, errors accumulate rapidly and destroy the encoded quantum information.
Quantum information processing offers a fundamentally different model of computation, with the potential to expand the range of problems that can be computed, simulated, and optimized beyond the practical reach of classical machines~\cite{shor1999polynomial, feynman2018simulating,peruzzo_variational_2014,McClean_2016,harrow2009quantum,QSVM,QPCA,montanaro2015quantum,PhysRevLett.125.260501,blank_quantum-enhanced_2021,DQA4BM}. Realizing this promise at scale, however, requires quantum processors that can operate reliably despite noise and imperfections. Error suppression~\cite{Viola1998two,Viola1999open,Uhrig2007UDD,Uys2009,Quiroz2011QDD,PhysRevA.98.013414,coote2025resource,jae2025measurement} and error mitigation~\cite{temme2017error,endo2018practical,9226505,Kurita2023synergeticquantum,Kim2023NatPhys,kim2023evidence,lee2023scalable,Cai2023} provide important tools for reducing the impact of such imperfections, particularly in near-term and early fault-tolerant settings. However, they do not guarantee a scalable route to arbitrarily long and reliable quantum computation. For that goal, active quantum error correction is essential: without repeated protection and correction, errors accumulate rapidly and eventually destroy the encoded quantum information.

Quantum error correction (QEC) provides the standard route toward scalable fault-tolerant quantum computation~\cite{shor1995scheme,548464,aharonov1997fault,rspa.1998.0166,KITAEV20032}. Instead of storing a logical qubit in a single physical qubit, QEC encodes logical information nonlocally into a larger entangled system of many physical qubits. Repeated stabilizer measurements are then used to extract information about errors without directly measuring, and thereby destroying, the protected logical state. When the physical error rate is below a threshold, increasing the size of the code can suppress the logical error rate, making long quantum computations possible in principle.

Topological quantum codes are among the leading candidates for implementing this idea in realistic hardware~\cite{KITAEV20032,10.1063/1.1499754,PhysRevA.86.032324,bombin2013introduction}. In these codes, logical information is protected by the global structure of the code rather than by any individual physical qubit. Local errors typically produce local syndrome signatures, while logical failures arise from error chains that become topologically nontrivial, for example by connecting appropriate boundaries of a surface-code patch in spacetime~\cite{bravyi1998quantum}. This geometric structure gives topological codes a natural compatibility with quantum hardware. In particular, the surface code has become a central architecture for fault-tolerant quantum computing because of its local stabilizer checks, relatively high threshold, and compatibility with two-dimensional processor layouts~\cite{PhysRevA.86.032324,Horsman_2012}.
The decoder is the classical inference engine in the QEC pipeline. Given a history of stabilizer measurement outcomes, it must infer the most likely logical error class or produce an appropriate recovery operation. In principle, optimal decoding corresponds to degenerate quantum maximum-likelihood decoding, in which one accounts for the total probability of stabilizer-equivalent error configurations. In practice, this problem is computationally intractable in general. Consequently, practical decoders must navigate a difficult trade-off: they should approach the accuracy of optimal decoding while remaining computationally efficient enough for deployment in realistic quantum computing systems.

Traditional decoding algorithms, such as minimum-weight perfect matching (MWPM), union-find decoders, belief-propagation-based methods, and tensor-network decoders, have played a central role in the development of QEC. These methods exploit mathematical structure in the code and noise model, and in several settings they provide highly effective and efficient decoding. Nevertheless, realistic quantum devices introduce complications that are difficult to capture with simple hand-designed assumptions. Noise can be biased, correlated, non-Pauli, nonstationary, or device-specific; readout may provide analog or soft information beyond binary syndrome bits; and fault-tolerant operations may generate syndrome patterns that differ substantially from those of simple memory experiments. These features motivate decoders that can adapt directly to data rather than relying entirely on an analytically specified noise model.

Machine learning (ML) fits naturally into this setting because quantum error decoding is, at the level of the decoder, a problem of processing classical data. The input is a high-dimensional syndrome history with spatial structure inherited from the code geometry and temporal structure inherited from repeated stabilizer measurements. The output may be a logical error class, a recovery operation, a set of marginal error probabilities, or a probability distribution over possible errors. In this sense, decoding can be viewed as a supervised classification problem, a probabilistic generative modeling problem, or a sequential decision-making problem, depending on how the task is formulated. Neural networks for instance can learn nonlinear mappings from these syndrome histories to logical corrections, or can model conditional distributions over errors and logical classes. Their flexibility makes them attractive for capturing correlations, using soft measurement information, adapting to experimental data, and integrating architectural priors such as locality, recurrence, graph structure, and attention.

This viewpoint is useful for both communities that this chapter aims to connect. For quantum computing experts, ML provides a flexible toolkit for building decoders that can learn from simulated or experimental data, adapt to hardware-specific noise, and incorporate architectural priors such as locality, recurrence, graph structure, attention, and state-space dynamics. For ML and AI experts, QEC decoding provides a scientifically meaningful learning problem with unusual constraints: extremely low target error rates, strong spatiotemporal structure, distribution shifts between simulation and hardware, severe latency requirements, and performance metrics tied to fault-tolerant computation rather than ordinary classification accuracy.

Recent progress suggests that ML-based decoders are no longer merely proof-of-concept models. Neural decoders based on recurrent networks, graph neural networks, Transformer-based architectures, and state-space models have been applied to surface codes, color codes, and other stabilizer-code families. Some of these models have demonstrated strong performance on experimental processor data, while others emphasize scalability, transfer across code distances, or real-time throughput. At the same time, these results also clarify the central difficulty: a practical ML decoder must simultaneously achieve high accuracy, scale to large code distances, generalize beyond its training distribution, and run fast enough for real-time QEC.

This chapter surveys machine learning approaches to quantum error decoding, with emphasis on architectural principles, practical performance, and real-time considerations. We focus primarily on topological stabilizer codes, especially surface-code decoding, while also discussing ideas that extend to color codes and other stabilizer-code settings. Rather than presenting ML decoders as a single algorithmic family, we organize the discussion around the key design choices that determine their behavior: the learning formulation, the output representation, the neural architecture, the training distribution, and the deployment constraints.

The chapter is organized as follows. In Sec.~\ref{sec:2}, we formulate decoding as a learning problem and review major ML paradigms, including discriminative, generative, and reinforcement-learning approaches. We discuss how the input space, label space, data-generating distribution, and irreducible ambiguity of syndrome data shape the learning problem. In Sec.~\ref{sec:3}, we introduce neural network building blocks and architectural principles relevant to QEC decoding, including recurrent neural networks, graph neural networks, transformer-based models, and state-space architectures such as Mamba. In Sec.~\ref{sec:4}, we review recent progress and benchmarks in neural decoding, focusing primarily on quantum memory experiments, which remain the most common and systematically studied setting for evaluating decoder performance. Section~\ref{sec:5} discusses real-time decoding and hardware considerations, including throughput, latency, streaming inference, and the role of accelerated classical hardware. Finally, Sec.~\ref{sec:6} summarizes open challenges and future directions, including data efficiency, generalization across code distances and geometries, robustness to hardware drift, and the integration of learned decoders into scalable fault-tolerant architectures.

% Throughout the chapter, we focus primarily on topological stabilizer codes, which provide a natural and widely studied setting for neural decoding, while many of the conceptual insights extend more broadly to other quantum error-correcting codes and noise models.

\section{Machine Learning Paradigms for Quantum Error Decoding}
\label{sec:2}

Machine learning enters quantum error decoding when decoding is viewed as an inference problem over classical data generated by a quantum error-correction experiment. Repeated stabilizer measurements convert the effects of physical errors into a syndrome history, from which the decoder must infer the logical effect of the underlying error. The structure of this inference problem is determined by the quantum code and hardware: the code geometry fixes the syndrome space, the stabilizer checks determine which error information is accessible, the noise process determines which error configurations are likely, and stabilizer equivalence defines what it means for two corrections to be physically identical. The learning problem is then to choose how to represent the syndrome data, what target the model should predict, and how the model should be trained and evaluated.

This perspective separates several choices that are often implicit in discussions of decoding. A learned decoder may be trained to predict a logical class directly, estimate marginal error probabilities, model a conditional distribution over errors, or construct a correction through a sequence of decisions. These choices lead naturally to different learning paradigms. Discriminative models learn a direct map from syndrome data to logical labels or recovery-related outputs. Generative models aim to represent a probability distribution over errors or logical cosets conditioned on the syndrome. Reinforcement-learning approaches instead formulate decoding as a sequential decision-making problem, in which an agent builds a correction by acting on syndrome information.

These paradigms are not mutually exclusive. Modern neural decoders often combine ideas from several of them: a model may be trained with a discriminative loss while producing calibrated probabilistic outputs, use auxiliary prediction tasks to improve training, or incorporate architectural components that impose locality, graph structure, recurrence, or attention. The appropriate formulation depends on the code family, the noise model, the available training data, and the intended deployment setting. In particular, real-time decoding places constraints not only on logical error rate but also on inference latency and scalability with code distance.

This section introduces these learning paradigms and clarifies how they relate to the structure of quantum error correction. We first formulate decoding as a statistical learning problem, beginning with the optimal decoding rule and then discussing supervised classification, training-data generation, and evaluation metrics. We then compare discriminative and generative approaches, emphasizing how the choice of output representation shapes their respective strengths and limitations. Finally, we briefly discuss reinforcement-learning formulations, which view decoding as a sequential correction problem.

\subsection{Decoding as a learning problem}
\label{subsec:2.1}

\subsubsection{The optimal decoder}
\label{subsubsec:2.1.1}

An $[\![n, k, d]\!]$ stabilizer code encodes $k$~logical qubits into $n$~physical qubits with distance~$d$~\cite{gottesman1997stabilizer}. In each QEC cycle, $m = n - k$ stabilizer generators are measured, producing a syndrome $\mathbf{s} \in \{0,1\}^m$. Any Pauli error~$E$ can then be decomposed as $E = T_\mathbf{s} \cdot L \cdot S$, where $T_\mathbf{s}$ is the pure error determined by the syndrome, $L \in \mathcal{L}$ is a logical operator, and $S \in \mathcal{S}$ is a stabilizer. Since errors that differ only by a stabilizer element are physically equivalent---the degeneracy central to quantum codes---the decoder's task is to identify the logical coset, not the physical error itself.

Given a syndrome~$\mathbf{s}$ and a noise model $\mathrm{Pr}(E)$, a conceptually simple strategy is non-degenerate quantum maximum-likelihood decoding (QMLD)~\cite{hsieh2011np, kuo2012hardness}, which selects the single most likely error consistent with the syndrome:
\begin{equation}
\hat{E}_\mathrm{QMLD} = T_\mathbf{s} \cdot \arg\max_{L \in \mathcal{L},\, S \in \mathcal{S}} \mathrm{Pr}(L, S \mid T_\mathbf{s}),
\label{eq:qmld}
\end{equation}
where $T_\mathbf{s}$ is the pure error uniquely determined by the syndrome. QMLD, however, disregards degeneracy and may select an error from a coset whose total probability is smaller than that of a competing coset. The optimal strategy is degenerate quantum maximum-likelihood decoding (DQMLD), which selects the logical coset with the largest total probability:
\begin{equation}
\hat{L}^* = \arg\max_{L \in \mathcal{L}} P(L \,|\, \mathbf{s}), \qquad P(L \,|\, \mathbf{s}) = \sum_{S \in \mathcal{S}} \mathrm{Pr}(L, S \mid T_\mathbf{s}).
\label{eq:dqmld}
\end{equation}
DQMLD achieves the lowest possible logical error rate for any code and noise model, and accounting for degeneracy can change the threshold itself~\cite{7097029}.

Both decoding problems are computationally intractable in general. QMLD is NP-hard and DQMLD is \#P-complete~\cite{7097029, hsieh2011np, kuo2012hardness}. This fundamental intractability necessitates learned decoders that can approximate DQMLD accuracy while maintaining inference speeds competitive with efficient heuristics like MWPM or union-find~\cite{ref:MatchingComparison, higgott2022pymatching, delfosse2020linear, delfosse2021almost}.

\subsubsection{Decoding as supervised classification}
\label{subsubsec:2.1.2}

\begin{figure}[ht]
    \centering
    \includegraphics[width=0.89\textwidth]{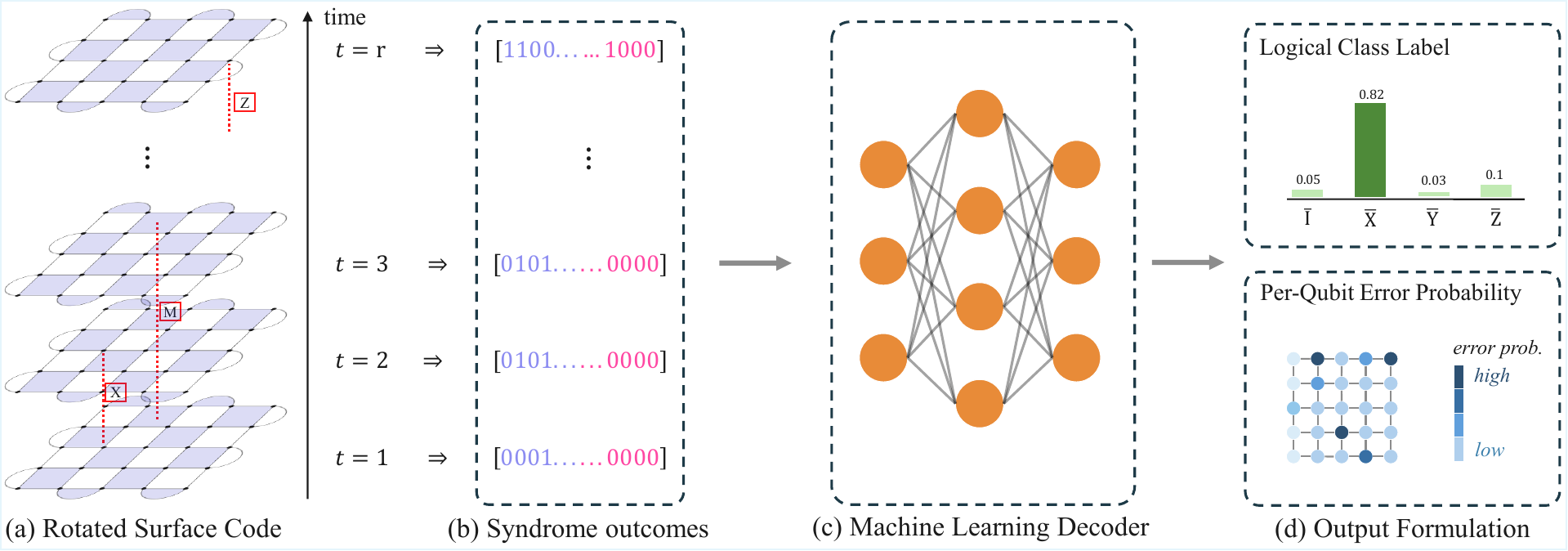}
    \caption{
    Overview of the ML-based decoding pipeline for stabilizer codes. (a) A rotated surface code undergoes repeated syndrome extraction over $r$ rounds; data qubits (black dots) are arranged on a lattice with $X$-type (white) and $Z$-type (blue) stabilizer measurements. (b) The measured syndrome outcomes at each round are represented as binary vectors and stacked across time steps to form the decoder input. (c) A machine-learning decoder---which may be instantiated as a feedforward network, CNN, GNN, recurrent network, or transformer---processes the syndrome history. (d) Two output formulations: a \emph{logical class label}, where the decoder predicts the most likely logical error class $\bar{L} \in \{\bar{I}, \bar{X}, \bar{Y}, \bar{Z}\}$ via softmax classification (top); or a \emph{per-qubit error probability} $P(E_i \,|\, \mathbf{s})$ for each data qubit, shown as colored nodes on the code lattice (bottom), where color intensity indicates error likelihood.
    }
    \label{fig:fig_1}
\end{figure}

Figure~\ref{fig:fig_1} illustrates the full decoding pipeline, from syndrome extraction to correction. A neural decoder $f : \mathcal{X} \to \mathcal{Y}$ is trained on syndrome--label pairs $\{(\mathbf{x}_i, y_i)\}$ to minimize the probability of misclassifying the logical coset, $R(f) = \mathrm{Pr}[f(\mathbf{x}) \neq y]$.

\textbf{Input space.}
Under code-capacity noise, where syndrome measurements are perfect, the input space is a single binary syndrome vector $\mathbf{s} \in \{0,1\}^m$. Circuit-level noise introduces measurement errors, requiring a syndrome \emph{history} over $r$ consecutive syndrome measurement rounds and a final round of data-qubit measurements (Fig.~\ref{fig:fig_1}a--b). The input may take the form of raw syndrome measurements (the stabilizer eigenvalues at each round) or detection events (changes between consecutive rounds). For a distance-$d$ surface code with $d^2 - 1$ stabilizers, $r$ syndrome measurement rounds, and one final data-qubit measurement round, the input has $(r+1)(d^2 - 1)$ binary entries. When the hardware exposes soft information---soft measurement posteriors derived from analog readout signals---the input space extends to $\mathcal{X} \subseteq \mathbb{R}^{n_\mathrm{input}}$, encoding richer constraints on the error distribution~\cite{varbanov2025neural, bausch2024learning}.

\textbf{Label space.}
For a code encoding $k$ logical qubits, the label set has $|\mathcal{Y}| = 4^k$ classes. For the standard $k=1$ case, this yields a four-class problem $\mathcal{Y} = \{\bar{I}, \bar{X}, \bar{Y}, \bar{Z}\}$ (Fig.~\ref{fig:fig_1}d, top). For CSS codes under independent $X/Z$ noise, the decoding problem factorizes into independent $\bar{X}$ and $\bar{Z}$ tasks, reducing each to a binary classification with $\mathcal{Y} = \{0,1\}$. An alternative formulation predicts per-qubit error probabilities $\hat{p} \in [0,1]^{2n}$, giving the marginal $X$ and $Z$ error probability on each physical qubit (Fig.~\ref{fig:fig_1}d, bottom)~\cite{krastanov2017deep}.

\textbf{Data-generating distribution and the irreducible error floor.}
A noise model parameterized by the physical error rate~$p$ assigns a probability $\mathrm{Pr}[E]$ to every error~$E$ on the $n$ data qubits. Each error determines both a syndrome $\mathbf{x}(E)$ and a logical class $L(E)$, inducing a joint distribution over inputs and labels,
\begin{equation}
  \mathcal{D}(\mathbf{x}, L)
    \;=\; \sum_{E:\,\mathbf{x}(E)=\mathbf{x},\;L(E)=L}
         \mathrm{Pr}[E].
  \label{eq:data_dist}
\end{equation}
This distribution is non-deterministic: errors that differ by a logical operator commute with every stabilizer and therefore produce the same syndrome, yet they belong to different logical cosets. A given syndrome therefore carries nonzero probability under multiple logical classes, and no decoder---regardless of its capacity---can eliminate this ambiguity. The optimal strategy is to return, for each syndrome, the most probable logical class; this is precisely DQMLD. Its error rate $R^{*}=R(f_{\mathrm{Bayes}})$, where $f_{\mathrm{Bayes}}(\mathbf{x})=\arg\max_{L}\,\mathrm{Pr}[L\mid\mathbf{x}]$, defines the irreducible floor against which every learned decoder is measured.

\textbf{Error decomposition.}
A neural decoder is trained on a finite sample $S = (\mathbf{x}_i, y_i)_{i=1}^N$ of $N$ pairs drawn i.i.d.\ from~$\mathcal{D}$. The population risk and the empirical risk on~$S$ are
\begin{align}
  R(f) := \mathrm{Pr}_{(\mathbf{x},y)\sim\mathcal{D}}\!\big[f(\mathbf{x}) \neq y\big],
  \qquad
  \widehat{R}_S(f) := \frac{1}{N}\sum_{i=1}^N \mathbf{1}\!\big\{f(\mathbf{x}_i) \neq y_i\big\}.
\end{align}
Within a hypothesis class~$\mathcal{F}$ (e.g., neural networks of a fixed architecture), let $f^*_\mathcal{F} \in \arg\min_{f\in\mathcal{F}} R(f)$ denote a best in-class predictor and $\widehat{f}_S \in \arg\min_{f\in\mathcal{F}} \widehat{R}_S(f)$ an empirical-risk minimizer. The training algorithm returns some $f_S \in \mathcal{F}$, which need not coincide with $\widehat{f}_S$; because $f_S$ depends on the random sample, $R(f_S)$ is itself a random variable, and bounds are stated in expectation or with high probability over~$S$.

The excess risk of $f_S$ over the DQMLD error rate~$R^*$ admits the four-term identity~\cite{berner2021modern}
\begin{align}
    R(f_S) - R^* \;=\;& \big[R(f_S) - \widehat{R}_S(f_S)\big] + \big[\widehat{R}_S(f_S) - \widehat{R}_S(f^*_\mathcal{F})\big] \nonumber\\
    & + \big[\widehat{R}_S(f^*_\mathcal{F}) - R(f^*_\mathcal{F})\big] + \big[R(f^*_\mathcal{F}) - R^*\big] \nonumber\\
    \;\le\;& \epsilon_\mathrm{opt} + 2\,\epsilon_\mathrm{gen} + \epsilon_\mathrm{app},
    \label{eq:error_decomp}
\end{align}
where $\epsilon_\mathrm{opt} := \widehat{R}_S(f_S) - \widehat{R}_S(\widehat{f}_S) \ge \widehat{R}_S(f_S) - \widehat{R}_S(f^*_\mathcal{F})$, $\epsilon_\mathrm{gen} := \sup_{f \in \mathcal{F}} |R(f) - \widehat{R}_S(f)|$, and $\epsilon_\mathrm{app} := R(f^*_\mathcal{F}) - R^*$.
The first and third bracketed terms are each bounded in absolute value by $\epsilon_\mathrm{gen}$, accounting for the factor of two; the second is at most $\epsilon_\mathrm{opt}$ because $\widehat{R}_S(\widehat{f}_S) \le \widehat{R}_S(f^*_\mathcal{F})$ by definition of $\widehat{f}_S$. Each named term targets a distinct source of suboptimality:
\begin{itemize}
    \item The \emph{approximation error}~$\epsilon_\mathrm{app}$ is the gap between the best in-class predictor $f^*_\mathcal{F}$ and the optimal decoder. The hypothesis class $\mathcal{F}$ is set by architecture (width, depth, connectivity); too small a class leaves $\epsilon_\mathrm{app} > 0$. Classical universal-approximation theorems guarantee that sufficiently wide networks can represent any continuous mapping on a compact input domain~\cite{lu2020universal}. Required capacity grows with code distance, syndrome dimensionality, and noise-model complexity.
    
    \item The \emph{generalization error}~$\epsilon_\mathrm{gen}$ is the worst-case gap between training-set and population risk over the hypothesis class. More training samples reduce $\epsilon_\mathrm{gen}$, while a richer $\mathcal{F}$ raises its worst-case bound. Standard regularization (weight decay, dropout, early stopping) and architectural inductive biases reduce the effective capacity searched by the optimizer, partially closing this gap. Required training-set size scales with the same factors, amplified by the capacity of $\mathcal{F}$.
    
    \item The \emph{optimization error}~$\epsilon_\mathrm{opt}$ is the gap between the predictor returned by training and the best one in $\mathcal{F}$ on the training data. Neural decoders are trained by SGD on a smooth proxy of the classification error (typically cross-entropy), since the classification error itself is not differentiable. The proxy's loss landscape is non-convex, so SGD can settle at a local rather than global minimum. Whether SGD finds a near-optimal predictor depends on the loss landscape and the chosen surrogate.
\end{itemize}

\begin{figure}[ht]
    \centering
    \includegraphics[width=0.88\textwidth]{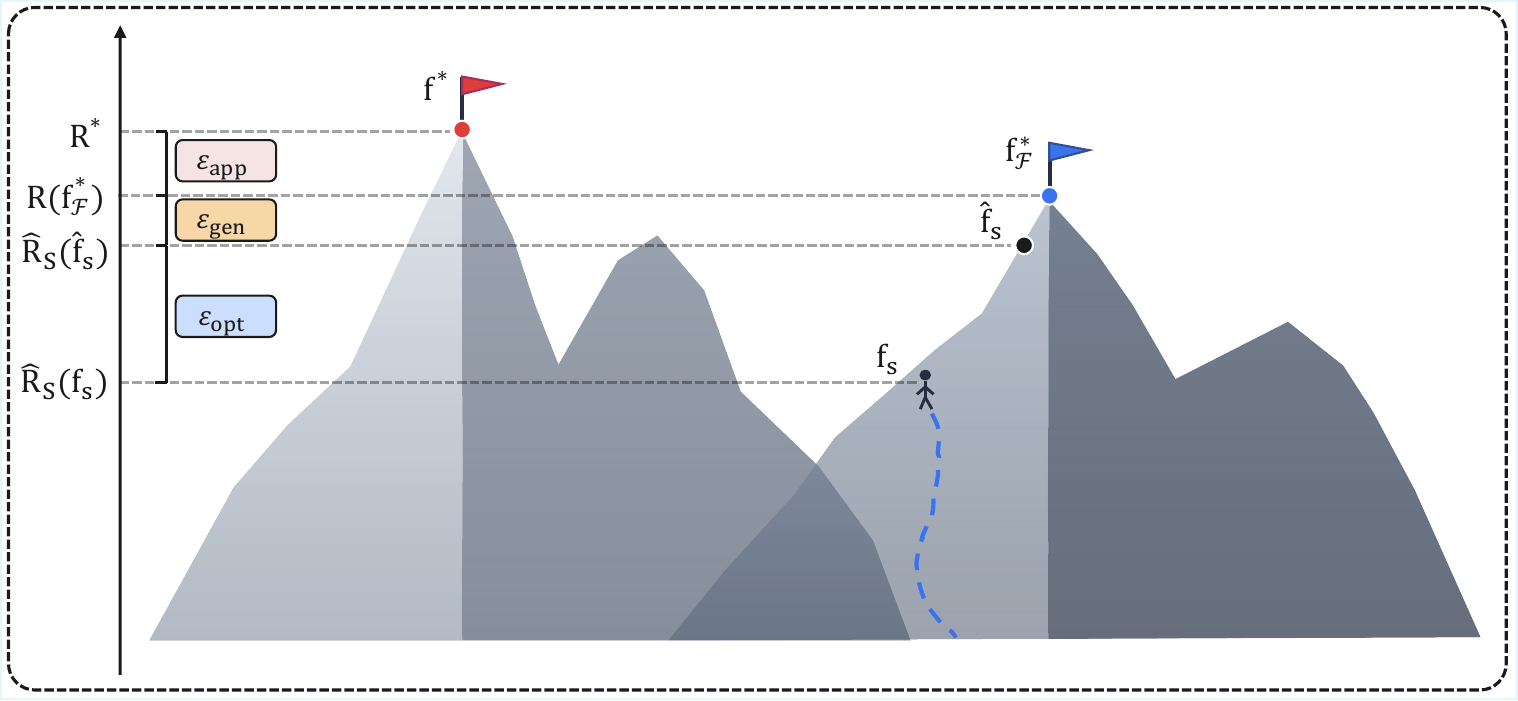}
    \caption{Visualization of the error decomposition.}
    \label{fig:fig_5}
\end{figure}
These three error sources map onto distinct categories of design decisions: architecture choice shapes $\epsilon_\mathrm{app}$, training-data strategy controls $\epsilon_\mathrm{gen}$, and the surrogate-loss and optimizer combination governs $\epsilon_\mathrm{opt}$. The next subsection examines training-data generation, importance sampling, and symmetry-based augmentation; Sec.~\ref{sec:3} surveys the architectural building blocks of neural decoders.

\subsubsection{Training data: generation, importance, and augmentation}

Training data are generated by sampling noisy stabilizer circuits using simulators such as Stim~\cite{gidney2021stim}. For each sample, a random error~$E_i$ is drawn from the noise model, the syndrome $\mathbf{x}_i = \mathbf{x}(E_i)$ is computed, and the logical class $y_i = L(E_i) \in \mathcal{Y}$ is read off from the simulator's tracking of the logical operator. Dataset sizes range from $\sim 10^6$ for small codes under code-capacity noise~\cite{Varsamopoulos_2018} to $\sim 2 \times 10^9$ samples per code distance for transformer-based decoders covering $d=3$ to $11$ under circuit-level noise~\cite{bausch2024learning}. Three practical challenges shape how these datasets are constructed: importance sampling, symmetry, and sim-to-real noise mismatch.

\textbf{Importance sampling.}
At low physical error rates, the vast majority of training syndromes correspond to low-weight errors that standard baseline decoders already handle correctly---making most of the training data uninformative. Training at an artificially elevated error rate increases the fraction of examples where the baseline fails but the optimal decoder succeeds. Peters~\cite{peters2025importance} formalized this trade-off between class imbalance and label noise, providing heuristic arguments for estimating the optimal amplification factor and demonstrating consistent improvements on a classical repetition-code toy problem (feedforward, convolutional, and transformer decoders) and on the $d=3$ surface code with a graph-based decoder.

\textbf{Symmetry.}
Codes defined on regular lattices possess discrete symmetries that permute stabilizers, map valid syndromes to valid syndromes, permute logical classes in a known way, and leave the noise distribution invariant under the appropriate noise model. Egorov~\emph{et al.}~\cite{egorov2023end} identify the toric code's automorphism group as generated by horizontal and vertical translations, $90^\circ$ rotations, horizontal flips, and a duality map exchanging $X$ and $Z$ between primal and dual sublattices. For surface codes the applicable symmetries depend on the exact lattice and boundary conditions~\cite{wagner2020symmetries}. Wagner~\emph{et al.}~\cite{wagner2020symmetries} map each syndrome to a unique translation representative before feeding it to a feedforward post-correction decoder; this enables improvements over MWPM at toric code lattice sizes $L=5,7$ where unaligned training fails to surpass MWPM at all. Egorov~\emph{et al.}~\cite{egorov2023end} instead embed translation equivariance directly into the architecture via twisted global average pooling on a wide-ResNet---the equivariant neural decoder (END)---leaving rotation, reflection, and duality equivariance to future work.

\textbf{Sim-to-real noise mismatch.}
A decoder trained on depolarizing noise will underperform on a device exhibiting biased, correlated, or circuit-level errors unless the training distribution matches the deployment target. The AlphaQubit decoder~\cite{bausch2024learning} addresses this with a two-stage pipeline: pretraining on billions of simulated samples from a circuit-level noise model, followed by fine-tuning on experimental data from the target processor. This sim-to-real transfer narrows the gap between idealized noise and device-specific error channels such as leakage and cross-talk.

\subsubsection{Evaluation metrics}

Decoder performance is conventionally reported as the logical error rate $p_L$ versus physical error rate~$p$ at several code distances~$d$. The key signature of a working decoder is that $p_L$ decreases with~$d$ at fixed~$p$ below the code's threshold $p_\mathrm{th}$. The error suppression ratio $\Lambda = p_L(d) / p_L(d+2)$~\cite{google2023suppressing, acharya2024quantum} quantifies how effectively each increment in distance suppresses errors. Beyond accuracy, practical evaluation requires reporting the decoding latency per syndrome and how both accuracy and latency scale with~$d$.

\subsection{Discriminative versus generative models}
\label{subsec:2.2}

The output representation of a neural decoder---whether a class label, per-qubit marginals, or a sample from the error distribution---determines both the data the decoder requires for training and the structural trade-offs it faces at inference. A discriminative decoder directly learns a mapping from syndromes to corrections: classification into $4^k$ logical cosets for a code encoding $k$ logical qubits, or per-qubit regression that predicts marginal error probabilities and delegates the final correction to a classical post-processing step~\cite{krastanov2017deep}. A generative decoder instead models the distribution over errors given the syndrome---either the conditional $P(E\,|\,\mathbf{s})$ directly~\cite{torlai2017neural} or a joint distribution over stabilizer configurations, logical operators, and syndromes from which coset probabilities can be extracted~\cite{cao2023qecgpt}. In either case, coset probabilities can be obtained by summing over stabilizer-equivalent errors---recovering DQMLD in principle, though the sum is exponentially costly unless the model's factorization permits efficient evaluation~\cite{cao2023qecgpt}---and training requires only syndrome--error pairs from the noise model rather than labeled corrections~\cite{cao2023qecgpt, cao2025generative}. The cost is that the network must represent a distribution over the full error space rather than a mapping to a compact label set. We examine each strategy and the trade-offs among them below.

\textbf{Discriminative decoders.}
Two output formulations coexist within the discriminative paradigm. The more common is \emph{classification}: the network learns $f_\theta\colon \mathbf{s} \mapsto \hat{L}$, predicting the logical coset via softmax over $|\mathcal{Y}|=4^k$ classes and trained with cross-entropy loss. Varsamopoulos~\emph{et al.}~\cite{Varsamopoulos_2018} established this template with a feedforward network on the surface code, testing noise models from code-capacity through phenomenological to circuit-level. Chamberland and Ronagh~\cite{chamberland2018deep} applied deep neural decoders to the rotated surface code as well as Steane and Knill error-correction protocols under full circuit-level noise; their network classifies the residual logical error left by a baseline lookup-table decoder and outputs an additional logical correction, rather than predicting the full physical recovery operator end-to-end. For CSS codes under independent $X/Z$ noise the problem factorizes into two binary classifications, but codes encoding $k\gg 1$ logical qubits face a $4^k$ label explosion that makes direct classification intractable. The leading discriminative decoder as of 2024, AlphaQubit~\cite{bausch2024learning}, retains a classification output and pairs it with a recurrent-transformer backbone; its architecture is discussed in Sec.~\ref{sec:3} and its benchmarks in Sec.~\ref{sec:4}.

The second formulation is \emph{per-qubit regression}: the network outputs marginal error probabilities $\hat{p} \in [0,1]^{2n}$, one for the $X$ and one for the $Z$ component on each physical qubit. Krastanov and Jiang~\cite{krastanov2017deep} introduced this approach on the toric code, where the shared hidden representation can implicitly exploit correlations between bit-flip and phase-flip channels that factored decoders treat independently. Because the output dimension scales as $2n$ rather than $4^k$, the regression formulation extends to high-rate codes without a combinatorial label explosion. The trade-off is that marginals alone do not specify a correction: a classical post-processor---syndrome-guided resampling in the original formulation~\cite{krastanov2017deep}---must assemble a valid recovery operator, and the overall decoder inherits whatever blind spots that post-processor has.

At inference, a discriminative decoder executes a single forward pass---deterministic and fixed-cost, with latency governed by the network size rather than by the code structure (hardware latency figures are discussed in Sec.~\ref{sec:5}). Training requires labeled syndrome samples, typically generated by a stabilizer-circuit simulator~\cite{gidney2021stim}; AlphaQubit~\cite{bausch2024learning} is pretrained on up to 2.5 billion simulated samples and fine-tuned on experimental data, using auxiliary prediction heads at intermediate QEC rounds and extracting multiple labels per training sequence to improve sample efficiency (Sec.~\ref{sec:4}). A second structural limitation is that standard discriminative training objectives encode no explicit model of the noise channel---though GNN-based decoders embed structural priors on the code through the graph topology: Lange~\emph{et al.}~\cite{lange2025data} construct a detector graph whose edges reflect space-time proximity between stabilizer events, while Ninkovic~\emph{et al.}~\cite{Ninkovic2024decoding} operate on the Tanner graph of quantum low-density parity-check (qLDPC) codes. In general, a decoder trained under depolarizing noise must be retrained or fine-tuned when deployed against biased or device-specific error distributions. On degeneracy: a classification decoder treats stabilizer-equivalent errors as a single label by construction but gains no probabilistic insight into the relative weight of different cosets, while the regression formulation produces soft marginals that a degeneracy-aware post-processor can exploit.

\textbf{Generative decoders.}
Generative decoders model the conditional distribution $P(E\,|\,\mathbf{s})$ rather than learning a direct syndrome-to-correction mapping. A sufficiently expressive generative model recovers the Bayes-optimal decoder, since the most probable logical coset can be read off from $P(E\,|\,\mathbf{s})$. The estimation cost discussed above is offset in practice by the ability to train directly from the noise model---sampling syndrome--error pairs without requiring a reference decoder to produce correction labels.

Torlai and Melko~\cite{torlai2017neural} introduced the generative approach with a restricted Boltzmann machine (RBM) on the toric code under phase-flip noise. At inference the observed syndrome conditions the RBM and Gibbs sampling draws candidate errors from the learned distribution $P(E\,|\,\mathbf{s})$. Each decode requires multiple Gibbs steps to mix, making sampling slower than a single forward pass, but the learned distribution naturally captures the high degeneracy of error chains sharing a given syndrome.

The cost of iterative sampling motivated autoregressive alternatives. Cao~\emph{et al.}~\cite{cao2023qecgpt} proposed qecGPT, a transformer that factorizes the joint distribution over stabilizer generators, logical operators, and syndromes autoregressively, reducing decoding to $O(2k)$ forward passes---linear in the number of logical qubits $k$---rather than enumeration of $4^k$ cosets. The follow-up Generative Neural Decoder (GND)~\cite{cao2025generative} extended this paradigm---primarily using a masked autoregressive model (MADE), though a transformer variant is also evaluated---to larger codes, including $[\![72,12,6]\!]$ bivariate bicycle codes where $2\times 12=24$ forward passes replace the $4^{12}\approx 1.7\times 10^7$ coset enumerations required by exact maximum likelihood decoding. Both models train without supervision from the noise model alone---an advantage for code families where no efficient labeling oracle exists.

A different generative route replaces sequential token generation with parallel denoising: Liu~\emph{et al.}~\cite{liu2025decoding} applied masked diffusion models to qLDPC decoding, where a fully masked initial state is progressively unmasked over $T$ denoising steps conditioned on the syndrome, with each step assigning values to a subset of the remaining masked bits based on confidence scores. The number of steps $T$ can differ between training and inference, providing a tunable accuracy--latency knob: fewer steps yield faster but less accurate decoding, while the decoding time scales approximately linearly with $T$. Tian~\emph{et al.}~\cite{Tian2025} explored a hybrid path, using a quantum generative adversarial network (QGAN) as an auxiliary trainer that generates correction paths for a transformer-based surface-code decoder---an illustration that the boundary between paradigms is not always sharp, though QGAN-based decoding has so far been tested only under phenomenological noise on small surface codes.

Generative decoders have been validated under narrower noise settings than their discriminative counterparts, and no generative decoder has yet been tested on experimental hardware data; specific benchmarks and noise-model coverage are compared in Sec.~\ref{sec:4}.

% \textbf{Comparative analysis.} ...

\subsection{Reinforcement learning}
\label{subsec:2.3}
The error-decoding task can be formulated as a sequential decision problem: A quantum code interacts with a noisy \textit{environment}, and an \textit{agent}—the error decoder—chooses an \textit{action} $L$ corresponding to a local Pauli operation to correct errors $E$ that have occurred at each time step $t$. The \textit{state} of the code can be represented by the syndrome measurement $\mathbf{s}$, possibly together with a history of previous correction actions. After an action is applied, the syndrome is updated, and the agent receives a \textit{reward} signal $r$, which is designed to encourage corrections that return the code state to the correct logical subspace while penalizing long or harmful correction sequences. This process continues until a halting condition is reached.

If a future state depends only on a current state and action, rather than on the full history of previous states, the decoding process can be modeled as a Markov decision process (MDP). Specifically, this condition implies that probability distribution of a finite trajectory $\tau= (\mathbf{s}_1, L_1, r_1,\ldots,\mathbf{s}_T, L_T, r_T)$ can be written as
\begin{equation}
    p(\tau) = p(\mathbf{s}_1)\prod_{t=1}^{T-1}p(\mathbf{s}_{t+1},r_{t+1}|L_t,\mathbf{s}_t)\pi(L_t|\mathbf{s}_t),
\end{equation}
where the probability $\pi$ is called \textit{policy}, namely a rule or probability distribution that assigns actions to observed states. Reinforcement learning (RL) provides a general framework for learning an optimal policy that maximizes cumulative reward $\mathbf{E}_{\tau\sim p(\tau)}[\sum_{t=1}^T \gamma^{t-1}r_t]$ with a discount factor $0<\gamma\le1$. In this formulation, decoding is not treated as a one-shot classification problem from syndrome to error decoding, but as a policy-learning problem in which the agent learns how to navigate the syndrome space toward successful logical recovery. The RL decoder learns a policy solely from reward signals obtained through trial-and-error interactions with the noisy code environment, which distinguishes this approach from the discriminative and generative models.

Sweke \emph{et al.} introduced RL formulations of QEC decoding in a fault-tolerant setting, explicitly allowing for faulty syndrome measurements~\cite{sweke2020reinforcement}. Using deep Q-learning, they trained decoding agents for the $d=5$ surface code under bit-flip and depolarizing noise. This work showed that decoding can be treated as adaptive interaction with a noisy code environment, rather than as a fully hand-designed syndrome-to-correction rule.

Andreasson \emph{et al.} then applied deep RL to bit-flip decoding in Kitaev’s toric code~\cite{andreasson2019quantum}. The syndrome configuration was used as the state, and each action corresponded to moving a defect to a neighboring site. By exploiting translational invariance, they simplified the state representation and trained a convolutional Q-network that achieved performance close to MWPM for code distances up to $d=7$. This provided a demonstration that a self-trained RL agent can rediscover a near-optimal decoding strategy in a minimal topological-code benchmark.

A closely related direction was developed by Colomer \emph{et al.}, emphasizing reward design in deep-RL decoding for the toric code~\cite{colomer2020reinforcement}. Instead of guiding the agent toward short correction paths, this work rewarded the decoder according to whether the logical state was preserved. With this logically motivated reward, the learned decoder achieved near-optimal performance around the toric-code threshold of approximately $11\%$, and the policy resembled MWPM. This strengthened the idea that RL can learn directly from the operational success criterion of QEC.

%: logical-state preservation.

Fitzek \emph{et al.} extended this line of work to toric-code decoding under depolarizing noise~\cite{Fitzek2020PRR}. Since depolarizing noise contains $X$, $Y$, and $Z$ errors, $Y$ errors induce correlations between bit-flip and phase-flip components. Their deep Q-learning decoder learned action-values for local Pauli corrections and outperformed standard MWPM for depolarizing noise, with improved success rates and thresholds up to $d\le9$. This showed that RL decoding can do more than reproduce MWPM-like behavior: it can exploit error correlations missed by separate $X$- and $Z$-type decoding.

Most existing RL-based QEC decoders are value-based, specifically relying on deep Q-learning, because early toric- and surface-code benchmarks use discrete local actions for which action-values $Q(s,a)$ can be estimated and greedily optimized. However, this choice is methodological rather than fundamental. A natural extension is policy-based RL, where the decoder directly learns a policy, potentially supporting stochastic recovery, higher-level actions, or recurrent/attention-based syndrome processing.

\section{Neural Network Building Blocks and Architectural Principles}
\label{sec:3}

This section introduces a small set of neural network architectures that have been most widely adopted as the backbone of contemporary neural decoders for quantum error correction. While many alternative architectures exist in the broader machine learning literature, the models discussed here capture the dominant design paradigms used in current decoding approaches. The first neural decoders were built around feedforward MLPs that classify flattened syndromes~\cite{Varsamopoulos_2018, krastanov2017deep, maskara2019advantages, overwater2022neural} and convolutional neural networks that exploit the lattice's spatial locality~\cite{meinerz2022scalable, gicev2023scalable, gicev2025fully}. In practice, modern neural decoders for quantum error correction often integrate multiple architectural components, combining, for example, recurrent structures for temporal memory with graph-based or attention-based modules for spatial and long-range correlations. The architectures reviewed here---RNNs, GNNs, transformers, and Mamba---represent commonly used building blocks that can be composed in different ways to balance expressivity, scalability, and real-time performance.

% Explicitly discuss some of the routinely used ML models like:
% \begin{itemize}
%     \item RNN
%     \item Graph NN
%     \item Transformer
%     \item Mamba
% \end{itemize}

\begin{figure}[t]
    \centering
    \includegraphics[width=0.89\textwidth]{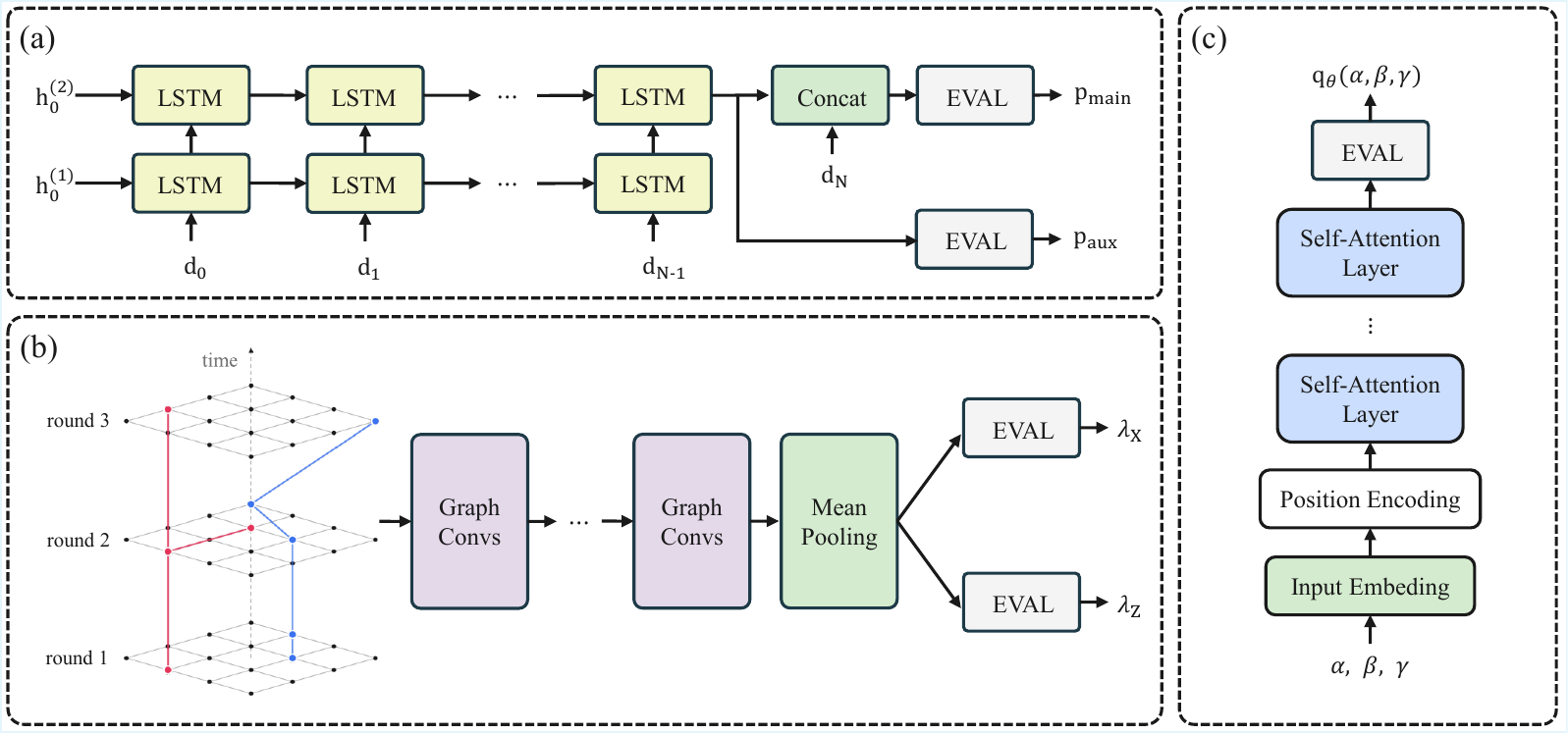}
    \caption{
    Three canonical neural decoder architectures.
    \textbf{(a)}~Two stacked LSTM layers process defects $d_0, d_1, \ldots, d_{N-1}$ sequentially. The auxiliary head receives the recurrent output alone and predicts $p_\mathrm{aux}$; the main head additionally ingests the final defects $d_N$ from data-qubit readout and predicts $p_\mathrm{main}$.
    \textbf{(b)}~Detector error graph spanning three measurement rounds; red nodes are $X$-type and blue nodes are $Z$-type detection events, with edges weighted by shared fault probability. Successive graph-convolution layers propagate local features along the space-time graph; mean pooling collapses all node embeddings into a fixed-size vector, and two evaluation heads output $\lambda_X$ and $\lambda_Z$.
    \textbf{(c)}~qecGPT causal Transformer. The joint configuration $(\boldsymbol{\alpha}, \boldsymbol{\beta}, \boldsymbol{\gamma})$ of stabilizer generators, logical operators, and syndrome is first projected via an input embedding, augmented with positional encoding, then passed through $N_l = 2$--$3$ stacked self-attention layers; the EVAL head outputs the joint distribution $q_\theta(\boldsymbol{\alpha},\boldsymbol{\beta},\boldsymbol{\gamma})$.
    }
    \label{fig:nn_building_blocks}
\end{figure}

\subsection{Recurrent Neural Networks}
\label{subsec:3.1}

Fault-tolerant syndrome extraction requires multiple rounds of stabilizer measurement. Data-qubit errors persist across subsequent syndrome rounds, while ancilla measurement errors affect only a single round's outcome. A single syndrome snapshot cannot resolve this ambiguity. The decoder must instead process the full syndrome history $\mathbf{s}^{(1)}, \ldots, \mathbf{s}^{(r)}$. Recurrent neural networks (RNNs) map this syndrome history to a correction by maintaining a hidden state $\mathbf{h}^{(t)}$ updated at each round:
\begin{equation}
\mathbf{h}^{(t)} = \sigma\!\bigl(W_h \mathbf{h}^{(t-1)} + W_x \mathbf{s}^{(t)} + \mathbf{b}\bigr),
\label{eq:rnn}
\end{equation}
where $W_h$ and $W_x$ are learned weight matrices and $\sigma$ is a nonlinear activation. The hidden state $\mathbf{h}^{(t)}$ compresses the syndrome history up to round $t$ into a fixed-size vector. The decoder outputs its correction from $\mathbf{h}^{(r)}$ at the final round.

Long short-term memory (LSTM) networks~\cite{hochreiter1997long}, adopted for QEC decoding~\cite{baireuther2018machine,baireuther2019neural}, extend the vanilla recurrence with a cell state governed by learned forget and input gates. Unlike feedforward networks, which require a fixed number of syndrome rounds and must be retrained whenever that count changes, LSTMs process syndrome rounds one at a time across an arbitrary number of rounds.

LSTM decoders have been demonstrated across different code families and extended to incorporate soft readout information from physical hardware. Baireuther~\emph{et al.}~\cite{baireuther2018machine} decoded the $d=3$ rotated surface code (Surface-17) using \emph{two separate networks}, each comprising two stacked LSTM layers ($N_L = 64$) followed by a fully connected evaluation layer. Network~1 receives all $T$ rounds of syndrome increments $\delta\vec{s}(t)$ from both $X$- and $Z$-stabilizer measurements and outputs the probability $p_1$ that bit-flip errors accumulated an odd parity over $T$ cycles. Network~2 receives only the final $T_0$ syndrome increments; its evaluation layer concatenates the last LSTM hidden state with the final syndrome increment
\begin{equation}
\delta\vec{f}(T) = \vec{f} - \vec{s}(T) \;\mathrm{mod}\;2,
\label{eq:baireuther_final_syndrome}
\end{equation}
computed from data-qubit readout, and outputs the probability $p_2$ that the readout requires an adjustment to the bulk estimate. The two probabilities combine as
\begin{equation}
p = p_1(1 - p_2) + p_2(1 - p_1),
\label{eq:baireuther_parity}
\end{equation}
giving the final parity correction probability for each logical basis independently. Baireuther~\emph{et al.}~\cite{baireuther2019neural} consolidated this two-network design into a single LSTM body with two output heads and applied it to the topological color code under circuit-level noise at $d = 3$, $5$, and $7$. The hidden-state dimension was scaled with distance ($N = 32$, $64$, $128$ for $d = 3$, $5$, $7$), and flag-qubit measurements were added as an additional input channel.

Varbanov~\emph{et al.}~\cite{varbanov2025neural} adopted the single-body two-head architecture of~\cite{baireuther2019neural} and extended it to incorporate soft readout information from transmon-qubit analog readout (Fig.~\ref{fig:nn_building_blocks}(a)). The shared recurrent body consists of two stacked LSTM layers with hidden-state dimension $N_L \in \{64, 96, 128\}$ for $d \in \{3, 5, 7\}$: the first LSTM outputs a hidden state per QEC round, the second LSTM processes that sequence and outputs only its final hidden state, and a ReLU activation follows. The auxiliary head receives the recurrent output alone and predicts $p_\mathrm{aux}$; the main head concatenates the recurrent output with the final defects $\{d_{a,N}\}$ inferred from data-qubit measurements and predicts $p_\mathrm{main}$. Training minimises the weighted binary cross-entropy
\begin{equation}
I = H(p_\mathrm{main},\, p_\mathrm{true}) + w_a\, H(p_\mathrm{aux},\, p_\mathrm{true}),
\label{eq:varbanov_loss}
\end{equation}
with $w_a = 0.5$, while evaluation uses $p_\mathrm{main}$ alone. Soft readout is incorporated by converting continuous I/Q measurement outcomes to per-measurement defect probabilities under a symmetric Gaussian noise model, preserving the confidence information that binary thresholding discards.

Two properties limit how well RNN decoders scale with code distance. First, the syndrome at each step enters as a flat vector with no encoding of stabilizer positions on the code lattice. Data-qubit errors correlate pairs of neighboring stabilizers in a geometry determined entirely by the code, but a serialized vector gives the network no indication of which indices are adjacent. Second, the hidden state $\mathbf{h}^{(t)}$ must be computed before $\mathbf{h}^{(t+1)}$, so the time axis cannot be parallelised. Graph-based (Sec.~\ref{subsec:3.2}) and attention-based (Sec.~\ref{subsec:3.3}) architectures address these two gaps directly. Recent decoders have combined RNN components with spatially-aware modules to exploit the strengths of both, as surveyed in Sec.~\ref{subsec:3.4}.

\subsection{Graph Neural Networks}
\label{subsec:3.2}

Stabilizer codes have inherent graph structure. Data qubits and stabilizer generators occupy opposing vertices of a bipartite Tanner graph, while repeated measurement rounds produce a space-time graph of detection events connected by shared fault mechanisms. Graph neural networks (GNNs; Fig.~\ref{fig:nn_building_blocks}(b)) exploit this structure through message passing, propagating local features along graph edges using shared weight matrices. This weight sharing fixes the trainable parameter count regardless of code distance or measurement rounds, so a single trained model generalizes across a range of code distances without retraining.

Two graph representations serve as the substrate for GNN decoding. The Tanner graph of an $[\![n,k,d]\!]$ stabilizer code is a bipartite graph with stabilizer generators on one side and data qubits on the other; edges connect each stabilizer to every qubit in its support. The detector error graph is constructed from the noise model, with nodes representing detection events and edges connecting event pairs that share a common fault mechanism.

A GNN operates on a graph $G = (V, E)$ in which node $i$ carries feature vector $\bar{X}_i$ and edge $(i,j)$ carries scalar weight $e_{ij}$. A single message-passing layer updates each node as:
\begin{equation}
    \bar{X}_i' = \sigma\!\bigl(W_1 \bar{X}_i + {\textstyle\sum_{j \in \mathcal{N}(i)}} e_{ji}\, W_2 \bar{X}_j\bigr),
    \label{eq:gnn_conv}
\end{equation}
where $W_1, W_2$ are learned weight matrices and $\sigma$ is a nonlinear activation. Lange~\emph{et al.}~\cite{lange2025data} instantiate this update on the detector error graph of the surface code. Each detection event is a node with feature vector $\bar{X}_i = (b_1,\,b_2,\,x,\,y,\,t)$, where $(b_1, b_2)$ encodes stabilizer type as a one-hot vector ($X$-type: $(1,0)$, $Z$-type: $(0,1)$), $(x,y)$ gives the lattice coordinates, and $t$ is the measurement round. Edge weights are set to the inverse Euclidean distance, $e_{ij} = \bigl((x_i-x_j)^2+(y_i-y_j)^2+(t_i-t_j)^2\bigr)^{-1/2}$. Seven convolutional layers feed into a mean-pooling operation that collapses all node embeddings into a single fixed-size graph embedding, after which two evaluation heads output $\lambda_X$ and $\lambda_Z$.

% Ninkovic~\emph{et al.}~\cite{Ninkovic2024decoding} and Maan and Paler~\cite{maan2025machine} operate on the Tanner graph rather than the detector error graph, targeting qLDPC codes. Ninkovic~\emph{et al.} define separate variable-to-check messages $m_{v \to c}^{k} = \mathrm{Message}^c(h_v^{k}, h_c^{k})$ and check-to-variable messages $m_{c \to v}^{k} = \mathrm{Message}^v(h_c^{k}, h_v^{k})$ across six GNN layers, with a single-layer gated recurrent unit (GRU)~\cite{} updating each node's hidden state; a per-qubit sigmoid head outputs marginal $X$- and $Z$-error probabilities for each data qubit~\cite{Ninkovic2024decoding}. Maan and Paler's Astra decoder~\cite{maan2025machine} parameterizes three local functions on the Tanner graph: a message MLP that computes directed messages from neighboring hidden states $(h_i^{t-1}, h_j^{t-1})$, a GRU that aggregates incoming messages with the node's syndrome input $x_j$ to produce updated hidden state $h_j^t$, and a readout MLP that maps the final hidden state to per-qubit error probabilities. Variable nodes are initialized to zero; check nodes are initialized with the observed syndrome values. All node states are updated simultaneously under a flooding schedule, and message passing runs for up to 100 iterations.

A second GNN approach uses the code's Tanner graph directly as the substrate, rather than the post-processed detector error graph. Hu~\emph{et al.}~\cite{hu2025efficient} construct an extended Tanner graph $G = (V_d \cup V_c \cup V_l, E)$ with three node types: data-qubit nodes $V_d$, check (stabilizer generator) nodes $V_c$, and logical-operator nodes $V_l$. Edges connect each check node to every data qubit in its stabilizer support, and each logical node to its defining qubit support. Syndrome values at each measurement round are embedded as check-node features, and a \emph{spatial extraction step} applies multiplicative message passing at each data node,
\begin{equation}
    v_n = \prod_{j \in \mathcal{N}(n)} \tanh\!\bigl(x_j\bigr),
    \label{eq:graphqec_mp}
\end{equation}
where $\mathcal{N}(n)$ denotes the check nodes adjacent to data node $n$ and $x_j$ are the check-node features. This product-of-tanh aggregation enforces parity-check constraint structure: the product at each data node captures the joint syndrome consistency of its neighboring stabilizers in a form isomorphic to their parity constraints~\cite{hu2025efficient}. The Tanner graph encodes only the stabilizer check matrix, not a specific noise model; the detector error graph, by contrast, incorporates fault probabilities into its edge weights at construction time and must be rebuilt when the noise distribution changes.

The Tanner graph connectivity differs substantially between code families, yet the message-passing update of Eq.~\eqref{eq:graphqec_mp} requires no modification for any of them. In the rotated surface code at distance $d$, bulk stabilizer generators have weight 4: each check node connects to four data-qubit nodes, and each bulk data qubit participates in two $X$-type and two $Z$-type checks. In the 2D triangular color code, bulk face stabilizers have weight 6 (six qubits per hexagonal face), and each data qubit participates in three $X$-type and three $Z$-type checks, giving degree 6 throughout the bulk. The 3-colorable trivalent lattice produces a Tanner graph that is strictly denser than the surface code case: where the surface code assigns each bulk data qubit to four stabilizer checks, the color code assigns it to six. The multiplicative update requires no architectural modification across code families: the product in Eq.~\eqref{eq:graphqec_mp} runs over degree-6 check neighborhoods for color code bulk nodes versus degree-4 for the surface code, with the model trained separately per code family. Hu~\emph{et al.}~\cite{hu2025efficient} verify this code-agnosticism directly, testing the same architecture on color codes from $[\![7,1,3]\!]$ through $[\![91,1,11]\!]$ under uniform depolarising noise by substituting the corresponding Tanner graph. The full architecture of GraphQEC---including temporal integration across syndrome rounds via linear attention and the logical-node readout---is described in Sec.~\ref{subsec:3.4}.

The Tanner-graph message-passing substrate extends to qLDPC codes with non-local, algebraically defined check structure~\cite{Ninkovic2024decoding,maan2025machine,hu2025efficient}; architectures and benchmarks for those decoders lie beyond this chapter's topological-code focus and are briefly discussed in Sec.~\ref{sec:6}.

\subsection{Transformers}
\label{subsec:3.3}

\subsubsection{The Self-Attention Mechanism}
\label{subsubsec:3.3.1}

In distance-$d$ topological codes, logical errors leave detection events only at the boundaries of their error support, potentially separated by graph distance $\sim d$; the decoder must correctly match these spatially distant events. Local architectures---CNNs, GNNs with few message-passing rounds---require $\sim d$ layers to propagate this information. Self-attention~\cite{vaswani2017attention} resolves this directly: every token pair interacts in a single layer, regardless of graph separation.

The core operation is scaled dot-product attention~\cite{vaswani2017attention}. Given query, key, and value matrices $\mathbf{Q}$, $\mathbf{K}$, $\mathbf{V} \in \mathbb{R}^{N \times d_k}$,
\begin{equation}
\mathrm{Attention}(\mathbf{Q}, \mathbf{K}, \mathbf{V}) = \mathrm{softmax}\!\left(\frac{\mathbf{Q}\mathbf{K}^\top}{\sqrt{d_k}}\right)\mathbf{V},
\label{eq:attention}
\end{equation}
where dividing by $\sqrt{d_k}$ prevents the softmax from saturating when $d_k$ is large. Multi-head attention runs $h$ such operations in parallel on lower-dimensional projections and concatenates the results:
\begin{equation}
\mathrm{MHA}(\mathbf{Q},\mathbf{K},\mathbf{V}) = \mathrm{Concat}(\mathrm{head}_1,\ldots,\mathrm{head}_h)\,\mathbf{W}^O,
\label{eq:mha}
\end{equation}
where $\mathrm{head}_i = \mathrm{Attention}(\mathbf{Q}\mathbf{W}_i^Q, \mathbf{K}\mathbf{W}_i^K, \mathbf{V}\mathbf{W}_i^V)$ and $d_k = d_\mathrm{model}/h$ per head. For a decoder that treats the full space-time syndrome as a flat token sequence---one token per (stabilizer, round) pair---the sequence length is $N = r \cdot m$ where $m \sim d^2$ is the number of stabilizers and $r$ is the number of measurement rounds. At $r = d$ this gives $N \sim d^3$, so each self-attention layer costs $O(d^6 d_k)$---prohibitive at large $d$ without temporal factorisation (Sec.~\ref{subsec:3.4}) or approximation.

Since syndrome tokens carry both spatial (lattice) and temporal (round) indices, 1D sequential positional encodings discard the 2D lattice structure; practical decoders instead inject geometry through two-dimensional rotary embeddings or learned spatial attention biases, as discussed in Sec.~\ref{subsec:3.4}.

\subsubsection{Transformer-Based Decoders}
\label{subsubsec:3.3.2}

Two patterns characterise how transformers have been deployed for topological-code decoding, differing in output representation and temporal processing strategy.

\textbf{Autoregressive generation.} As introduced in Sec.~\ref{subsec:2.2}, qecGPT~\cite{cao2023qecgpt} (Fig.~\ref{fig:nn_building_blocks}(c)) frames decoding as autoregressive generation over the triple $(\boldsymbol{\gamma}, \boldsymbol{\beta}, \boldsymbol{\alpha})$: syndrome $\boldsymbol{\gamma} \in \{0,1\}^m$ (commutation of $E$ with each stabilizer generator), logical operator configuration $\boldsymbol{\beta} \in \{0,1\}^{2k}$, and stabilizer generator configuration $\boldsymbol{\alpha} \in \{0,1\}^m$. A causal Transformer ($d_\mathrm{model} = 256$ for most tested codes, $512$ at $d=7$; $h = 4$ heads; $N_l = 2$--$3$ layers) factorises the joint distribution as
\[
q_\theta(\boldsymbol{\alpha},\boldsymbol{\beta},\boldsymbol{\gamma}) = q_\theta(\boldsymbol{\alpha} \mid \boldsymbol{\beta},\boldsymbol{\gamma})\cdot q_\theta(\boldsymbol{\beta} \mid \boldsymbol{\gamma})\cdot q_\theta(\boldsymbol{\gamma}),
\]
generating each of the $2k$ logical bits $\hat{\beta}_i = \arg\max\,q_\theta(\beta_i \mid \beta_{<i}, \boldsymbol{\gamma})$ in sequence. This reduces exact maximum-likelihood decoding from $4^k$-coset enumeration to $O(2k)$ autoregressive passes. Tested on rotated surface codes and toric codes under depolarizing noise, qecGPT surpasses MWPM; on the $[\![12,1,2]\!]$ 3D surface code---where MWPM does not apply directly---it closely matches exact maximum-likelihood decoding and significantly outperforms BPOSD. The $O(2k)$ gain grows with $k$, becoming substantial for codes with $k = 3, 5, 7$ logical qubits---obtained by removing stabilizers from a $d=3$ surface code patch---where exhaustive enumeration requires $4^k$ evaluations per syndrome. Accuracy benchmarks are reported in Sec.~\ref{sec:4}.

\textbf{Non-autoregressive discriminative.} AlphaQubit~1~\cite{bausch2024learning} takes the opposite approach: a single forward pass through a spatial transformer encoder that processes $m$ stabilizers per syndrome round, with recurrent modules managing temporal context across rounds. This temporal--spatial factorisation, which limits per-round attention cost to $O(m^2)$ rather than $O(r^2 m^2)$, is the defining architectural feature and is detailed in Sec.~\ref{subsec:3.4}.

QecGPT and AlphaQubit~1 show that attention can match or exceed MWPM on topological codes, but the quadratic scaling of full self-attention in the number of spacetime tokens becomes a real-time bottleneck at code distances relevant for fault tolerance. This tension between expressive global attention and hardware latency budgets motivates the factorized and hybrid designs examined next (Sec.~\ref{subsec:3.4}).

\subsection{State Space Models}
\label{subsec:3.4}

State space models (SSMs) parameterise sequence processing as a linear continuous-time dynamical system~\cite{hamilton1994state}. Similar to RNNs, they process sequences of length $L$ in $O(L)$ at inference. The linear recurrence admits parallel training that nonlinear recurrences cannot. This training takes the form of a global convolution when parameters are time-invariant and a parallel scan otherwise. The Structured State Space model (S4)~\cite{gu2021efficiently} is the canonical time-invariant instance, with HiPPO-initialised long-range memory~\cite{gu2020hippo}. Mamba~\cite{gu2023mamba} drops the time-invariance to gain content-aware mixing at linear-time cost.

The continuous-time dynamics are
\begin{equation}
\mathbf{h}'(t) = \mathbf{A}\,\mathbf{h}(t) + \mathbf{B}\,x(t), \qquad y(t) = \mathbf{C}\,\mathbf{h}(t),
\label{eq:ssm_cont}
\end{equation}
where $\mathbf{h}(t) \in \mathbb{R}^{d_h}$ is a latent state, $x(t)$ is a scalar input, and $\mathbf{A}$, $\mathbf{B}$, $\mathbf{C}$ are learned parameters. Discretising at step size $\Delta$ gives a recurrence
\begin{equation}
\mathbf{h}_k = \bar{\mathbf{A}}\,\mathbf{h}_{k-1} + \bar{\mathbf{B}}\,x_k, \qquad y_k = \mathbf{C}\,\mathbf{h}_k,
\label{eq:ssm_disc}
\end{equation}
with $(\bar{\mathbf{A}}, \bar{\mathbf{B}})$ depending on the discretisation rule. Mamba uses zero-order hold, giving $\bar{\mathbf{A}} = e^{\Delta \mathbf{A}}$ and $\bar{\mathbf{B}} = (\Delta \mathbf{A})^{-1}(e^{\Delta \mathbf{A}} - \mathbf{I})\,\Delta \mathbf{B}$. For S4, training computes the full output as an FFT-based convolution at near-linear cost in $L$~\cite{gu2021efficiently}; inference unrolls the recurrence at constant cost per token, suitable for streaming inference. Mamba makes $\mathbf{B}$, $\mathbf{C}$, and $\Delta$ learned functions of the current input $x_k$, lifting the time-invariance and enabling per-token modulation of the state-update rate. The model assigns small $\Delta$ to uninformative tokens, leaving $\bar{\mathbf{A}} \approx \mathbf{I}$ and passing the state through unchanged. Informative tokens receive large $\Delta$ and update the state sharply. The selective scan remains parallelisable via a hardware-aware parallel scan algorithm, replacing attention's $O(L^2)$ pairwise interactions with $O(L)$ work along the sequence dimension.

Lee~\emph{et al.}~\cite{lee2025scalable} apply this in surface-code decoding by replacing the Mixer Block of an AlphaQubit-style Syndrome Mixer with a Mamba module (Sec.~\ref{subsec:3.5}), reducing per-round mixing cost from $O(m^2)$ to $O(m)$ in $m = d^2 - 1$ stabilizers.

\subsection{Hybrid and Composed Architectures}
\label{subsec:3.5}

Recent neural decoders combine distinct building blocks for spatial and temporal mixing. The AlphaQubit family~\cite{bausch2024learning,senior2025scalablerealtimeneuraldecoder} and its Mamba-based variant~\cite{lee2025scalable} illustrate this pattern, pairing a per-round spatial mixer with a recurrent state across rounds.

\begin{figure}[t]
    \centering
    \includegraphics[width=0.89\textwidth]{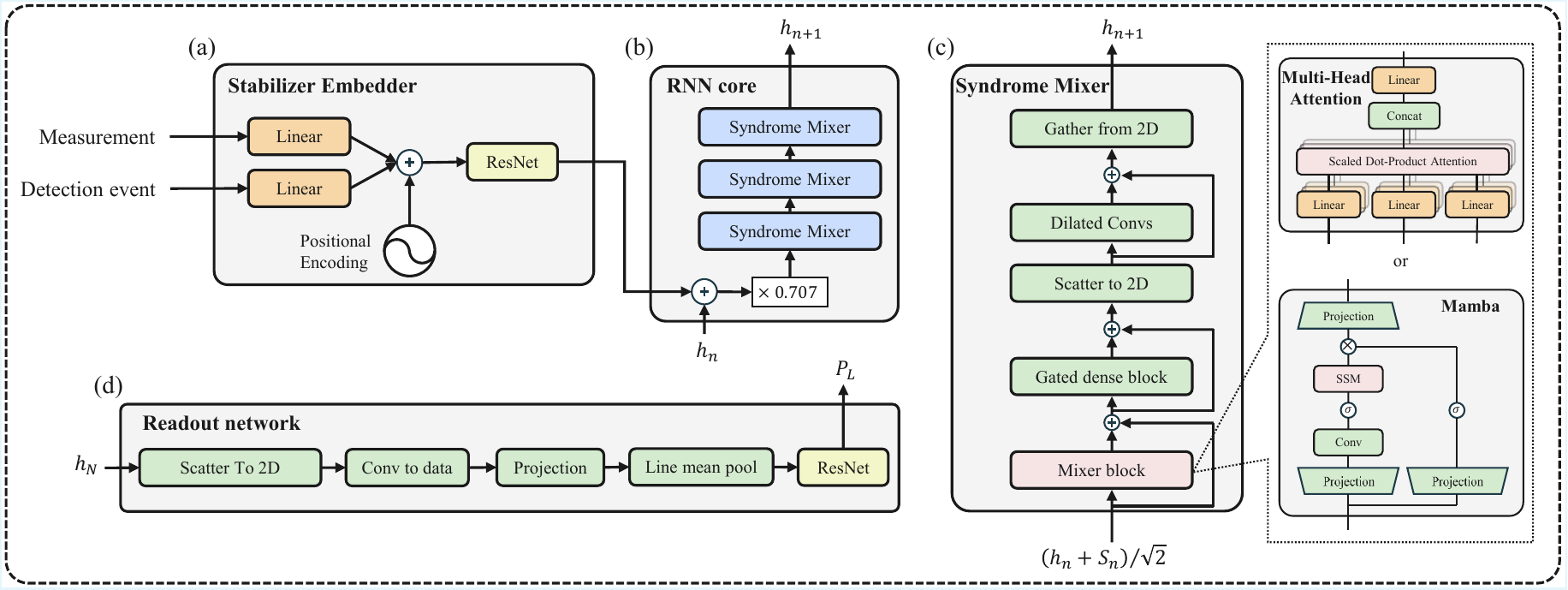}
    \includegraphics[width=0.89\textwidth]{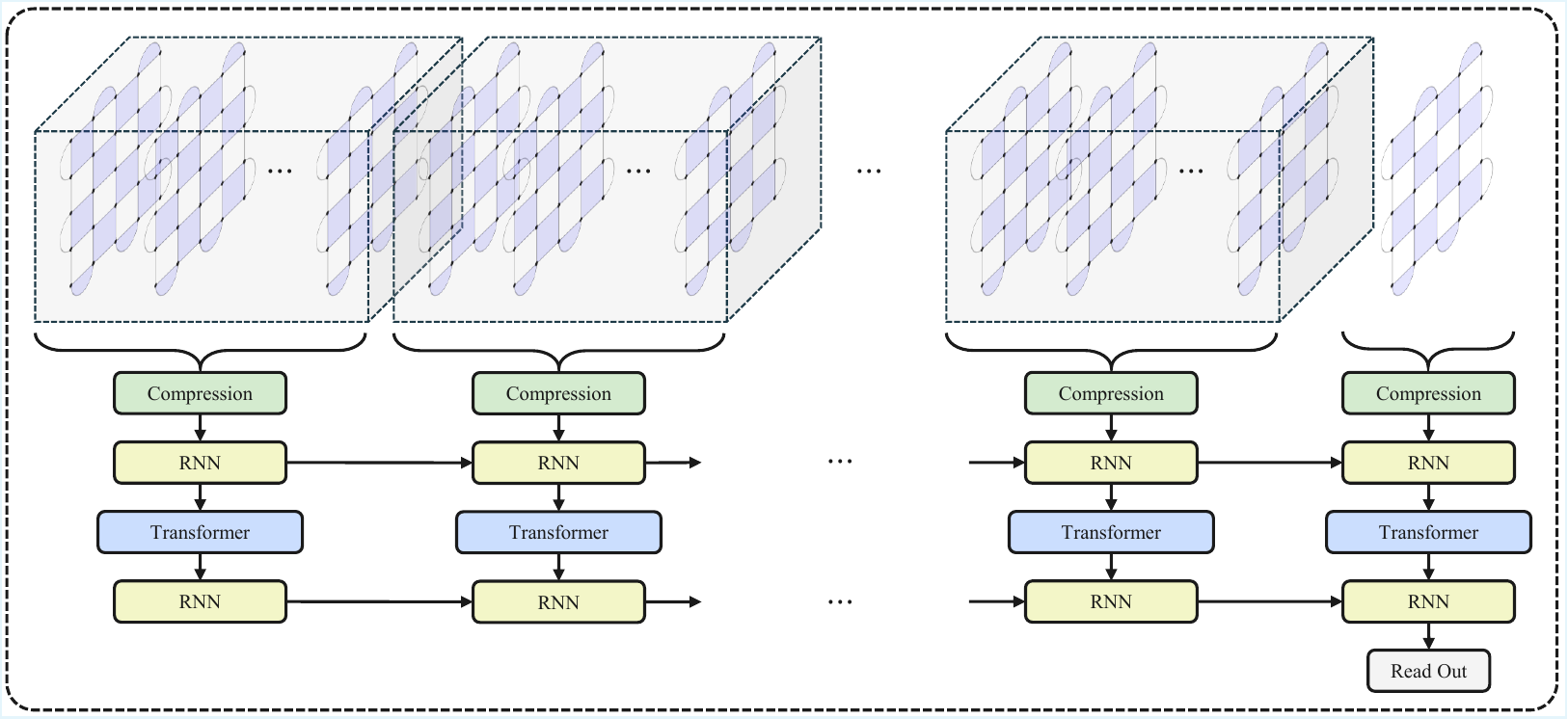}
    \caption{
    AlphaQubit family architectures. \textbf{Top:} AlphaQubit~1 (AQ1)~\cite{bausch2024learning} module pipeline. Adapted from Ref.~\cite{lee2023scalable}. \textbf{Bottom:} AlphaQubit~2 (AQ2)~\cite{senior2025scalablerealtimeneuraldecoder} with frame compression and alternating recurrent and transformer layers.
    }
    \label{fig:alphaqubit}
\end{figure}

AlphaQubit~1 (AQ1)~\cite{bausch2024learning} (Fig.~\ref{fig:alphaqubit}, top) consists of four modules: a \emph{Stabilizer Embedder} converts each round's per-stabilizer features into input embeddings; a \emph{Syndrome Mixer} combines multi-head self-attention and dilated 2D convolutions for within-round mixing; an \emph{RNN Core} integrates across rounds by repeatedly applying the Syndrome Mixer to a persistent per-stabilizer state; and a \emph{Readout Network} maps the final state to a logical-error probability.

Lee~\emph{et al.}~\cite{lee2025scalable} keep the AQ1 modules intact and replace only the Mixer Block's attention with a Mamba module (Sec.~\ref{subsec:3.4}). Each Mamba block branches into two parallel paths fused element-wise: an MLP path for local features, and a structured state-space scan that accumulates long-range context along the per-round stabilizer sequence at linear cost.

AlphaQubit~2 (AQ2)~\cite{senior2025scalablerealtimeneuraldecoder} (Fig.~\ref{fig:alphaqubit}, bottom) inherits the four-module structure of AQ1 but redesigns each module. Multiple syndrome cycles are compressed into single frames, and the AQ1 RNN Core is replaced by a lightweight recurrent cell. AQ2 drops all convolutions, encoding spatial structure with rotary position embeddings (RoPE)~\cite{su2024roformer} applied separately to physical-qubit $x$ and $y$ coordinates. The readout becomes cross-attention from per-observable queries to per-stabilizer tokens. A real-time variant (AQ2-RT) uses a Griffin-style element-wise gated recurrence.

\section{Recent Progress and Benchmarks in Neural Decoding}
\label{sec:4}

Most neural decoders are first evaluated on quantum-memory experiments, which provide a clean and standardized setting for assessing decoding performance. A logical state is prepared in a known basis, for example an eigenstate of logical $Z$ or $X$, followed by $r$ rounds of stabilizer measurement. At the end of the experiment, the data qubits are measured in a basis compatible with the prepared logical state. The decoding task is to infer whether the target logical observable has flipped during the experiment.

Let $\mathbf{s}^{(t)}$ denote the stabilizer measurement outcomes at round $t$. Under circuit-level noise, measurement errors are themselves noisy events, so the decoder is usually given detection events rather than raw stabilizer values,
\begin{equation}
    \mathbf{d}^{(t)}
    =
    \mathbf{s}^{(t)}
    \oplus
    \mathbf{s}^{(t-1)} ,
    \qquad
    t=1,\ldots,r ,
\end{equation}
together with final defects obtained by comparing the last syndrome round with the syndrome inferred from final data-qubit readout. The full decoder input is therefore a spacetime syndrome record,
$
    \mathbf{x}
    =
    \left(
    \mathbf{d}^{(1)},\ldots,\mathbf{d}^{(r)},\mathbf{d}^{(\mathrm{final})}
    \right).
$
For a single logical memory, the label can be written as a binary variable
$
    y \in \{0,1\},
$
where $y=1$ indicates that the measured final logical observable differs from the initially prepared logical value.

In a supervised neural decoder, a model $f_\theta$ maps the syndrome record to a logical-flip probability,
\begin{equation}
    f_\theta(\mathbf{x})
    =
    \Pr(y=1\,|\,\mathbf{x}) .
\end{equation}
The standard training objective is binary cross-entropy,
\begin{equation}
    \mathcal{L}(\theta)
    =
    -
    y\log f_\theta(\mathbf{x})
    -
    (1-y)\log\!\left(1-f_\theta(\mathbf{x})\right),
\end{equation}
or multiclass cross-entropy when the output is a full logical Pauli class. A decoding failure occurs when the predicted logical correction disagrees with the true logical flip. If the experiment contains $r$ QEC cycles, the measured memory-failure probability may be converted to an effective logical error per cycle under an independent-round approximation,
\begin{equation}
    p_{\mathrm{fail}}(r)
    =
    \frac{1-(1-2p_L)^{r}}{2}
    \approx
    r p_L ,
    \qquad
    p_L \ll 1 .
\end{equation}

This memory formulation is attractive for neural approaches because it gives a clean end-to-end learning problem: the input is the measured syndrome history and the target is the final logical outcome. It allows neural networks to learn from simulated stabilizer-circuit data, experimental samples, or a combination of both. It also allows the decoder to exploit information that is difficult to incorporate into simple analytic decoding assumptions, such as correlated faults, leakage indicators, analog readout information, or device-specific calibration structure. For this reason, memory experiments have become the standard benchmark for comparing neural decoder architectures, even though the same architectural ideas are also being extended to logical operations, high-rate codes, and real-time settings.

Before comparing benchmarks, it is useful to distinguish the noise and data regimes that appear throughout the literature. In \emph{code-capacity} or \emph{data-level} noise, Pauli errors are applied directly to data qubits and syndrome measurements are assumed perfect. This setting isolates the combinatorial decoding problem but omits measurement and circuit faults. In \emph{phenomenological} noise, data-qubit errors and measurement-bit errors are both included across repeated syndrome rounds, but the syndrome-extraction circuit itself is not explicitly modeled. In \emph{circuit-level} noise, faults are inserted after elementary operations such as gates, resets, measurements, and idling periods, producing the spacetime correlations and hook errors encountered in fault-tolerant circuits.

Many recent benchmarks use stabilizer-circuit simulators such as Stim~\cite{gidney2021stim}. A common circuit-level benchmark is the superconducting-inspired SI1000 model, which assigns nonuniform depolarizing, measurement, reset, and idle errors intended to approximate superconducting QEC cycles~\cite{gidney2021stim,senior2025scalablerealtimeneuraldecoder}. Other studies use hardware-inspired or device-calibrated simulations. For example, AlphaQubit uses SI1000 pretraining and Pauli+ simulation data with leakage and full I/Q readout information for larger-distance studies~\cite{bausch2024learning}. Real experimental data are also increasingly important. The Google Sycamore surface-code memory dataset, released with the corresponding experiment, contains executed circuits, measured samples, and decoder outputs, and has become a standard testbed for data-driven decoders~\cite{google2023suppressing}.

Table~\ref{tab:neural_decoding_benchmarks} summarizes representative neural-decoding benchmarks, ordered chronologically. Since memory experiments remain the dominant benchmark in this area, many entries use the memory setup described above. The table is nevertheless intended as a broader survey of neural decoding progress, including works that emphasize generative modeling, graph structure, scalable architectures, experimental data, and real-time-oriented deployment. The entries should not be read as a direct ranking by logical error rate, since the benchmarks differ in code family, distance, noise model, number of syndrome rounds, input information, training distribution, and baseline decoder.

\begingroup
\scriptsize
\setlength{\tabcolsep}{2.4pt}
\renewcommand{\arraystretch}{1.15}

% \begin{longtable}{L{0.045\textwidth} L{0.135\textwidth} L{0.175\textwidth} L{0.215\textwidth} L{0.345\textwidth}}

\begin{xltabular}{0.89\textwidth}{
@{}
L{0.06\textwidth}
L{0.16\textwidth}
L{0.17\textwidth}
L{0.20\textwidth}
X
@{}
}
\caption{
Representative ML-based decoding benchmarks for topological quantum codes, ordered chronologically. The table is not intended to be exhaustive. It emphasizes works that introduced influential neural-decoding formulations, architectural ideas, experimental-data benchmarks, scalability studies, or extensions beyond quantum memory. Strictly qLDPC-only and non-topological-code studies are omitted.
}
\label{tab:neural_decoding_benchmarks}\\

\toprule
\textbf{Year} &
\textbf{Work} &
\textbf{Code and benchmark} &
\textbf{Noise model or data} &
\textbf{ML formulation and main contribution} \\
\midrule
\endfirsthead

\caption[]{Representative ML-based decoding benchmarks for topological quantum codes, continued.}\\
\toprule
\textbf{Year} &
\textbf{Work} &
\textbf{Code and benchmark} &
\textbf{Noise model or data} &
\textbf{ML formulation and main contribution} \\
\midrule
\endhead

\midrule
\multicolumn{5}{r}{\emph{Continued on next page}}\\
\endfoot

\bottomrule
\endlastfoot

2017 &
Torlai and Melko~\cite{torlai2017neural} &
Toric-code memory &
Code-capacity phase-flip noise with perfect syndrome extraction &
Restricted Boltzmann machine generative decoder. Introduced generative neural decoding for topological codes by learning an error distribution conditioned on the syndrome. \\

2017 &
Krastanov and Jiang~\cite{krastanov2017deep} &
Stabilizer-code memories, with toric-code demonstrations &
Code-capacity Pauli noise, including depolarizing noise in toric-code benchmarks &
Feedforward neural decoder predicting per-qubit error probabilities. Shifted the output from a global logical class to local error marginals, followed by classical post-processing. \\

2017 &
Varsamopoulos \emph{et al.}~\cite{Varsamopoulos_2018} &
Small-distance surface-code memory &
Code-capacity, phenomenological, and circuit-level surface-code noise models &
Feedforward neural decoder. Provided an early systematic study of neural surface-code decoding, including network size, dataset size, noise model, and inference-time trade-offs. \\

2018 &
Baireuther \emph{et al.}~\cite{baireuther2018machine} &
Rotated surface code, Surface-17 memory &
Density-matrix simulations of repeated-round Surface-17 experiments with correlated $X$ and $Z$ errors &
LSTM recurrent decoder. Showed that recurrent networks can process syndrome histories directly and exploit temporal correlations in repeated-round surface-code memory experiments. \\

2018 &
Chamberland and Ronagh~\cite{chamberland2018deep} &
Surface-code, Steane-code, and Knill-style near-term error-correction experiments &
Full circuit-level noise in small-distance fault-tolerant circuits &
Deep neural residual decoder after a baseline correction stage. Demonstrated neural decoding in realistic near-term circuit-level settings by learning residual logical corrections. \\

2019 &
Baireuther \emph{et al.}~\cite{baireuther2019neural} &
Topological color-code memory &
Circuit-level color-code noise with flag-qubit information &
LSTM recurrent decoder with auxiliary output structure. Extended recurrent neural decoding from surface-code memories to color-code memories with circuit-level noise. \\

2019 &
Maskara \emph{et al.}~\cite{maskara2019advantages} &
Toric- and triangular color-code memories &
Several topological-code noise models, including spatially correlated errors &
Versatile feedforward neural decoder. Emphasized adaptability to structured and correlated noise beyond standard matching assumptions. \\

2019 &
Andreasson \emph{et al.}~\cite{andreasson2019quantum} & Toric-code memory & Bit flip noise with perfect syndrome extraction& Simplified representation of syndrome measurements using translational invariance. Outperformed MWPM using a convolutional Q-network.\\

2020 &
Colomer \emph{et al.}~\cite{colomer2020reinforcement} & Toric-code memory & Uncorrelated bit or phase flip noise with perfect syndrome extraction & Emphasized reward design in reinforcement-learning-based decoding by rewarding the deep Q-learning agent for preserving the logical state, rather than encouraging short correction paths.\\

2020 &
Fitzek \emph{et al.}~\cite{Fitzek2020PRR} & Toric-code memory & Depolarizing noise and perfect syndrome extraction & Outperformed MWPM with improved success rates and thresholds up to $d\le 9$ using Q-learning decoder. \\

2020 &
Sweke \emph{et al.}~\cite{sweke2020reinforcement} &
Topological-code decoding for fault-tolerant settings &
Simulated repeated-round syndrome data with measurement noise &
Reinforcement-learning decoder. Framed decoding as a sequential decision problem in which an agent constructs corrections from syndrome information. \\

2020 &
Ni~\cite{ni2020neural} &
Large-distance two-dimensional toric-code memory &
Code-capacity bit-flip noise with perfect syndrome extraction &
Scalable neural decoder inspired by renormalization. Addressed the distance-scaling problem by targeting toric-code instances much larger than early proof-of-principle neural decoders. \\

2022 &
Meinerz \emph{et al.}~\cite{meinerz2022scalable} &
Surface-code memory at large scale &
Simulated surface-code noise for scalable topological-code benchmarks &
Scalable local neural decoder. Used local processing to improve distance scaling relative to monolithic neural decoders. \\

2023 &
Egorov \emph{et al.}~\cite{egorov2023end} &
Toric-code memory &
Simulated toric-code noise distributions for symmetry-aware decoding &
Equivariant neural decoder. Built lattice symmetries directly into the architecture, reducing reliance on data augmentation and improving sample efficiency. \\

2023 &
Cao \emph{et al.}~\cite{cao2023qecgpt} &
Surface-code and toric-code memory benchmarks &
Depolarizing and correlated-noise models on stabilizer-code benchmarks &
Autoregressive Transformer generative decoder. Recast decoding as autoregressive modeling of syndromes, stabilizers, and logical operators, reducing logical-coset evaluation from exhaustive enumeration to sequential generation. \\

2024 &
AlphaQubit~\cite{bausch2024learning} &
Rotated surface-code memory, including Sycamore experimental data &
Google Sycamore distance-three and distance-five memory data. Larger-distance studies use SI1000 pretraining and Pauli+ simulation data with leakage and I/Q readout information &
Recurrent Transformer-based logical decoder. Established a high-accuracy sim-to-real neural-decoding pipeline using simulated pretraining and experimental-data fine-tuning. \\

2025 &
Varbanov \emph{et al.}~\cite{varbanov2025neural} &
Small-distance rotated surface-code memories with transmon readout &
Circuit-level simulations, Google Sycamore experimental memory data, and synthetic soft-readout data &
LSTM decoder using hard or soft readout information. Showed that recurrent neural decoders can exploit analog readout information in near-term surface-code memory settings. \\

2025 &
Lange \emph{et al.}~\cite{lange2025data} &
Surface-code memory represented as a detector graph &
Circuit-level surface-code simulations, with tests on Google experimental repetition-code data &
Graph neural network on detection-event graphs. Encoded detector geometry and local fault connectivity directly into the neural architecture. \\

2025 &
Hu \emph{et al.}~\cite{hu2025efficient} &
Surface-code, color-code, and stabilizer-code memory benchmarks &
Uniform depolarizing simulations and Sycamore experimental surface-code data &
GraphQEC architecture using stabilizer-graph structure and sequence modeling. Demonstrated a more code-agnostic neural architecture across topological and stabilizer-code families. \\

2025 &
Zhou \emph{et al.}~\cite{zhou2025learning} &
Surface-code logical circuits with single-qubit and entangling logical gates &
Mirror-symmetric random Clifford circuits and deeper logical-circuit benchmarks &
Multi-Core Circuit Decoder. Moved beyond memory decoding by learning reusable modules for logical operations and gate-induced correlations. \\

2025 &
Ataides \emph{et al.}~\cite{ataides2025neural} &
Universal-algorithm decoding with surface and color codes &
Circuit-level algorithmic workloads with realistic noise features, including loss-resolving readout in relevant settings &
Modular attention-based neural decoder. Extended neural decoding from memories to algorithmic workloads by learning gate-induced correlations and decoding relevant logical observables. \\

2025 &
Lee \emph{et al.}~\cite{lee2025scalable} &
Surface-code memory and simulated real-time scenarios &
Sycamore hardware memory data and simulated real-time decoding scenarios &
Mamba-based state-space neural decoder. Replaced attention with state-space sequence modeling to improve the speed-accuracy trade-off for real-time-oriented surface-code decoding. \\

2025 &
AlphaQubit~2~\cite{senior2025scalablerealtimeneuraldecoder} &
Surface- and color-code memory benchmarks at larger distances &
Stim-generated SI1000 circuit-level data, with hardware-data fine-tuning in reported studies &
Scalable spatiotemporal neural decoder with real-time-oriented variants. Connected high-accuracy topological-code decoding with throughput, model size, and accelerator-friendly inference. \\

2026 &
Zhang \emph{et al.}~\cite{zhang2026learning} &
Surface-code real-time decoding with temporal window decomposition &
Simulated surface-code syndrome histories for window-decomposed decoding &
Self-coordinating neural-window decoder. Changed the output from a single global logical bit to window-local logical contributions, making neural decoding more compatible with parallel-window execution. \\

2026 &
Gu \emph{et al.}~\cite{gu2026scalable} &
Surface-code memory benchmarks, with geometric-code extensions &
Stim-generated circuit-level depolarizing memory experiments and data-level depolarizing studies &
Geometry-aware convolutional neural decoder. Used local convolutional structure and code geometry to obtain a hardware-friendly decoder architecture with favorable throughput characteristics. \\

\end{xltabular}
\endgroup

Several trends emerge from these benchmarks. Early neural decoders established that supervised, generative, and reinforcement-learning methods can learn nontrivial decoding maps on small topological codes. Later work introduced recurrence, graph structure, attention, equivariance, and state-space dynamics to better match the spacetime structure of syndrome data. More recent studies have shifted from proof-of-principle memory decoding toward scalability, experimental-data adaptation, qLDPC codes, logical circuits, and hardware-aware inference. The field has therefore moved from asking whether neural networks can decode at all to asking which architectural and systems choices make learned decoders useful in realistic fault-tolerant workflows.
%Recent work has also begun to explore hybrid learned--algorithmic decoding pipelines for surface-code memory experiments. In particular, AI-based pre-decoders use neural networks to perform local space--time corrections and reduce syndrome density before passing the residual syndrome to a conventional global decoder such as PyMatching~\cite{chamberland2026fast}. This approach is attractive because it preserves the role of established algorithmic decoders while using learned local processing to improve runtime and, in some regimes, logical error rate. Although current demonstrations focus primarily on memory settings, the locality and block-wise structure of pre-decoding make it a promising direction for future real-time decoding in larger surface-code architectures and lattice-surgery-based logical operations.

% \subsection{Logical Operations}

\section{Real-Time Decoding and Hardware Considerations}
\label{sec:5}

Real-time decoding should be distinguished from the decoding task encountered in quantum-memory experiments or, more generally, in fault-tolerant protocols restricted to Clifford operations. In a memory experiment, the full syndrome record can often be decoded after the experiment has finished, because the decoder output is needed primarily to interpret the final logical measurement or to determine the final Pauli frame. Likewise, Clifford gates map Pauli operators to Pauli operators under conjugation, so an unresolved Pauli frame can be propagated classically through the circuit without changing the structure of the computation. In this regime, slow decoding delays the availability of the final interpreted result, but it does not necessarily force the quantum processor itself to pause.

The situation changes in the presence of adaptive non-Clifford operations. There, decoder output is no longer needed only at the end of the experiment; it may be required during the computation in order to determine subsequent classically controlled corrections or measurement interpretations. As a result, decoding must be treated not merely as a post-processing task, but as part of the online control loop of the fault-tolerant processor. The remainder of this section is organized as follows. Section~\ref{subsec:5.1} introduces the exponential backlog problem and explains why insufficient decoding throughput can become a severe bottleneck in computations with many adaptive non-Clifford steps. Section~\ref{subsec:5.2} reviews sliding-window and parallel-window techniques proposed to mitigate this problem. Section~\ref{subsec:5.3} then turns to neural architectures for real-time decoding, with emphasis on the architectural features needed for low-latency inference and high sustained throughput. Finally, Section~\ref{subsec:5.4} discusses hardware-dependent timing budgets and deployment constraints across different qubit platforms.

\subsection{Exponential Backlog Problem}
\label{subsec:5.1}

The distinction between memory and real-time decoding becomes operationally important once the computation contains adaptive non-Clifford primitives, such as magic-state injection, $T$-gate teleportation, or classically controlled logical measurements. In such circuits, the current logical Pauli frame can influence which correction should be applied next, or how a subsequent measurement outcome should be interpreted. It is still not necessary to physically apply every Pauli correction after each QEC cycle, since Pauli-frame tracking remains the preferred implementation. However, the relevant logical-frame information must be available before the next adaptive feed-forward step. Real-time decoding therefore requires the classical control system to maintain a sufficiently up-to-date logical frame for the computation to proceed without an unbounded waiting time~\cite{PhysRevA.86.032324,terhal2015quantum}.

This distinction motivates two performance requirements. The first is \emph{sustained throughput}: the decoder must process syndrome data at least as fast as the quantum device generates it. The second is \emph{reaction latency}: when an adaptive logical operation requires a decoded decision, the relevant frame information must be returned within the allowed classical-control window. A decoder may have high average throughput while still returning a final decision many QEC cycles after the corresponding measurements. Conversely, a decoder may have low latency on small instances but insufficient aggregate throughput when many logical patches are decoded simultaneously. Practical real-time decoding must satisfy both constraints at the system level.

Let $\tau_{\mathrm{QEC}}$ denote the duration of one QEC cycle and let $l$ be the number of syndrome bits generated per cycle by a distance-$d$ code block. The syndrome-generation rate is
$
    r_{\mathrm{gen}} = {l}/\tau_{\mathrm{QEC}}.
$
If the classical decoding infrastructure processes syndrome information at rate $r_{\mathrm{proc}}$, then the basic throughput condition is
$
    r_{\mathrm{proc}} \ge r_{\mathrm{gen}} .
$
Equivalently, after accounting for parallel workers, buffering, communication, and window overlap, the average amount of classical work generated by one QEC cycle must be completed within one QEC cycle. For surface-code patches, $l$ grows as $O(d^2)$ per round, so the throughput requirement becomes increasingly stringent as the code distance and the number of simultaneously active logical patches increase.

The consequence of violating this condition is the decoding-backlog problem. Following Ref.~\cite{terhal2015quantum}, suppose that syndrome information is generated at rate $r_{\mathrm{gen}}$ but processed at a slower rate $r_{\mathrm{proc}}$, and define
$
    f = {r_{\mathrm{gen}}}/{r_{\mathrm{proc}}} > 1 .
$
If an unresolved delay $\Delta_{\mathrm{gen}}$ produces a backlog
$
    D_1 = r_{\mathrm{gen}}\Delta_{\mathrm{gen}},
$
then processing this backlog requires time
$
    \Delta_{\mathrm{proc}}
    =
    {D_1}/{r_{\mathrm{proc}}}
    =
    f\,\Delta_{\mathrm{gen}} .
$
During that time, new syndrome data are generated, producing a next unresolved block of size
$
    D_2 = r_{\mathrm{gen}}\Delta_{\mathrm{proc}} = fD_1 .
$
Repeated over $k$ adaptive synchronization points, the unresolved data volume grows as
$
    D_k = f^k D_1 .
$
Thus, if the decoder cannot keep up with the syndrome stream, repeated non-Clifford feed-forward steps can induce a geometric growth in the waiting time. By contrast, in a memory experiment or a Clifford-only computation, slow decoding primarily delays classical post-processing rather than repeatedly stalling the quantum computation.

The end-to-end reaction time for an adaptive decision is platform dependent, but it can be schematically decomposed as
$
    \tau_{\mathrm{react}}
    \approx
    \tau_{\mathrm{readout}}
    +
    \tau_{\mathrm{det}}
    +
    \tau_{\mathrm{dec}}
    +
    \tau_{\mathrm{frame}}
    +
    \tau_{\mathrm{ctrl}},
$ where the terms denote readout classification, detection-event generation, decoding, Pauli-frame or logical-frame update, and classical-control feedback. These stages can often be pipelined, and different platforms distribute the cost differently. The important point is that neural-network inference time is only one component of the real-time decoding stack.

\subsection{Sliding-Window and Parallel-Window Techniques}
\label{subsec:5.2}

A natural way to reduce decoding latency and memory footprint is to avoid decoding the entire spacetime syndrome history as a single global object. Sliding-window decoding partitions the incoming syndrome stream into temporal windows. Within each window, the decoder uses a bounded amount of syndrome history, commits corrections in a core region, and retains buffer regions to handle correlations that cross temporal boundaries. The window size must be large enough to preserve decoding accuracy, but it need not grow with the total duration of the computation. This makes sliding-window decoding better suited to online operation than full batch decoding.

\begin{figure}[!htbp]
    \centering
    \includegraphics[width=0.89\textwidth]{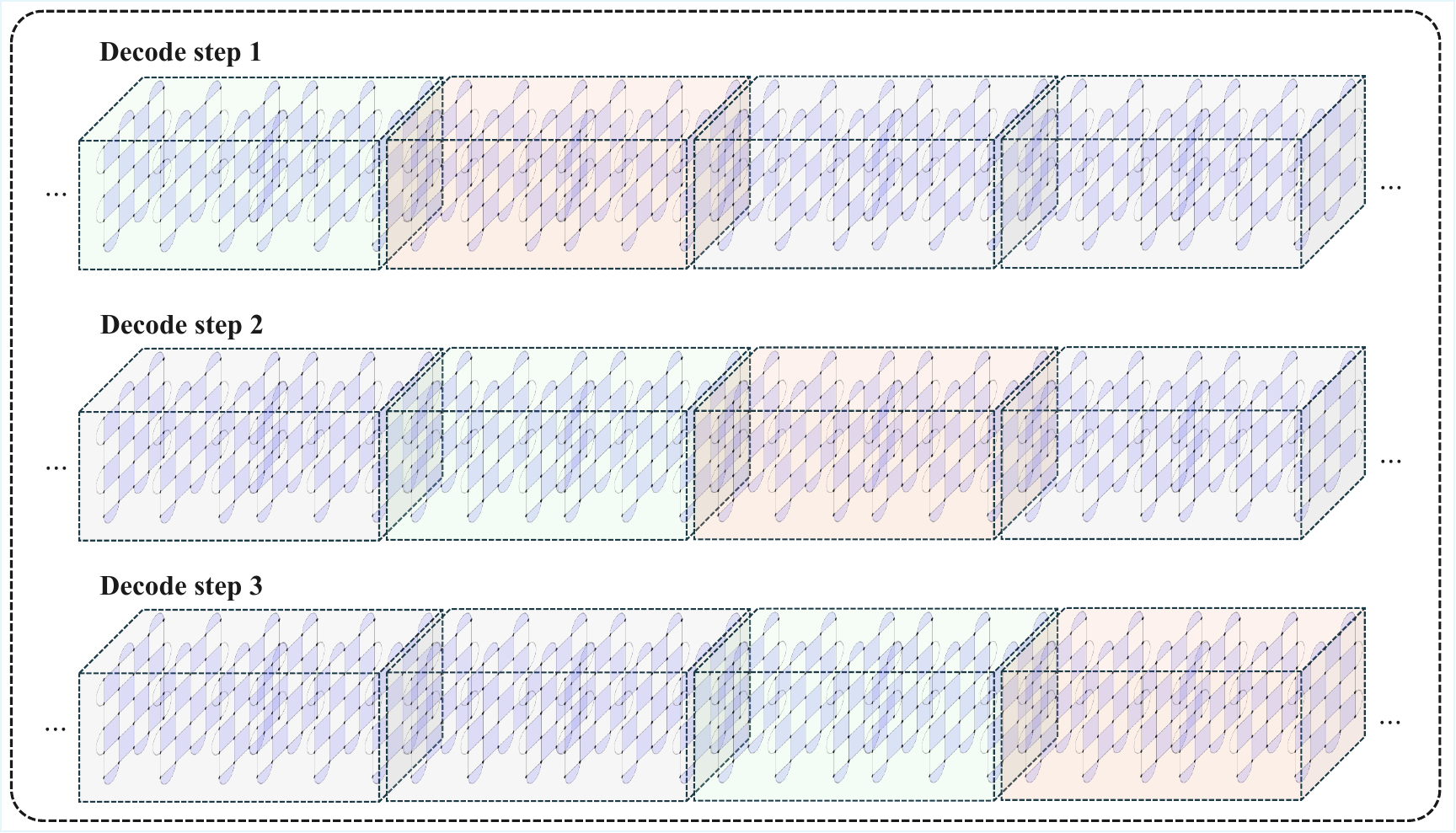}
    \vspace{0.5em}
    \includegraphics[width=0.89\textwidth]{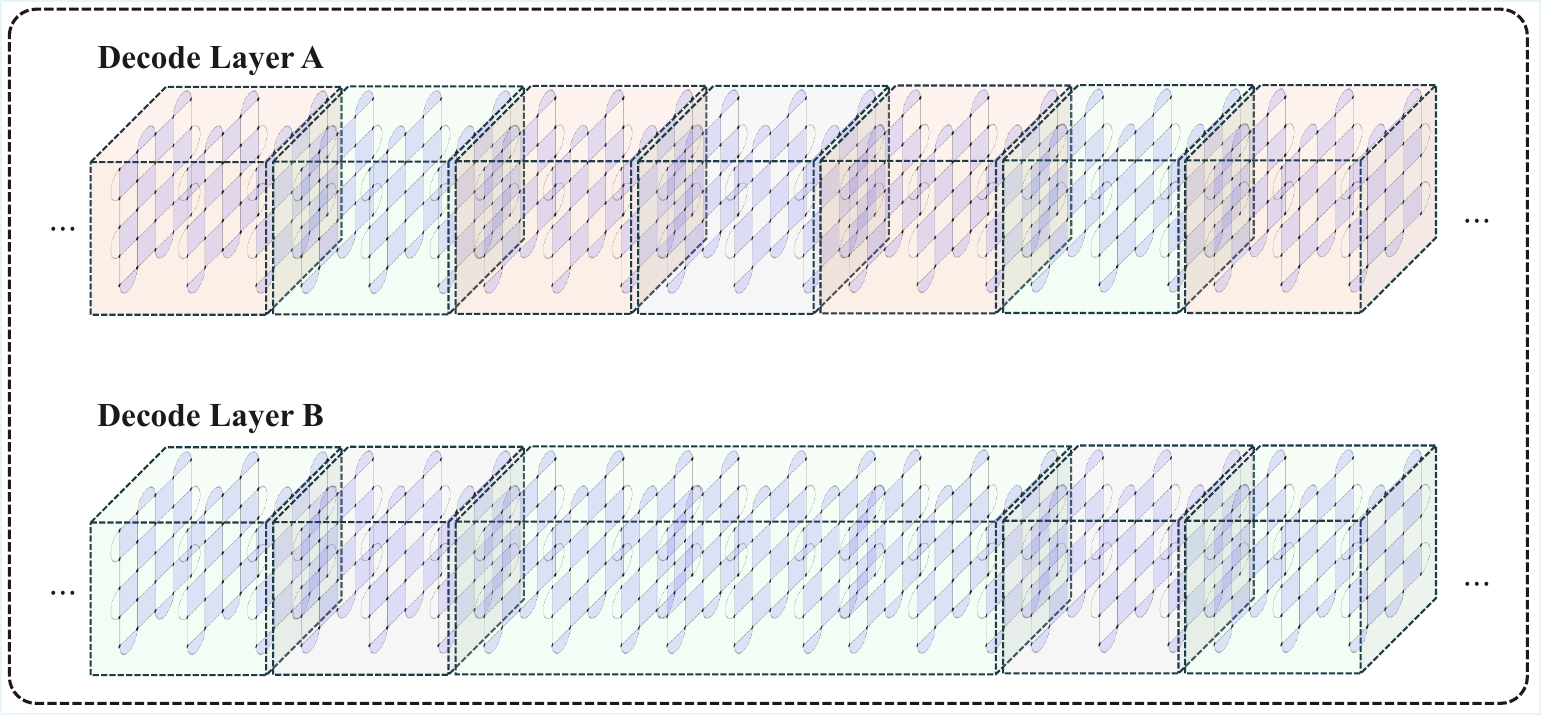}
    \caption{
    Sliding-window and parallel-window decoding. Green blocks indicate committed regions, where corrections are accepted and passed forward. Red blocks indicate buffer regions, which provide additional syndrome context but are not committed at that decoding step.
    \textbf{Top:} In sliding-window decoding, a temporal window is moved forward over the syndrome stream. At each decode step, the decoder commits the correction in the green core region and uses the red buffer region to account for error chains that extend across the window boundary. The buffer is retained until enough future syndrome information is available.
    \textbf{Bottom:} In parallel-window decoding, windows are organized into decoding layers. Within a layer, several non-overlapping windows can be decoded concurrently. Boundary information from the first layer is passed to a second layer, which resolves the regions left between the committed cores of the first-layer windows. In both cases, the purpose of the buffer is to localize the decoding problem while suppressing boundary-induced decoding errors.
    }
    \label{fig:window_decoding}
\end{figure}

Figure~\ref{fig:window_decoding} illustrates the two windowing regimes. In sliding-window decoding, a window is advanced along the time direction in successive decode steps. The green region denotes the committed part of the window, whose correction is accepted and passed to the Pauli-frame update. The red region denotes the buffer, which supplies additional syndrome context but is not yet committed. Parallel-window decoding uses the same commit-and-buffer principle, but arranges the windows into layers so that several non-overlapping decoding tasks can be executed at the same time.

In the two-layer construction of Ref.~\cite{skoric2023parallel}, the windows decoded within a given layer are chosen to be non-overlapping. In the first layer, multiple separated windows are decoded in parallel, and corrections are committed only in the high-confidence core regions. Tentative corrections that cross from a committed region into a buffer region generate artificial defects at the window boundaries. These artificial defects are then passed to the second layer, whose windows fill the unresolved regions between neighboring first-layer committed regions. Thus, parallelism is obtained not by independently committing all adjacent temporal windows at once, but by arranging the decoding into layers of mutually non-overlapping windows together with a reconciliation mechanism between layers. In the MWPM simulations of Ref.~\cite{skoric2023parallel}, choosing the committed and buffer regions to have widths comparable to the code distance led to no noticeable difference in logical error rate among global, sliding-window, and parallel-window decoding.

Related time-parallel sliding-window schemes use similar commit-and-buffer ideas, but differ in how temporal windows are arranged and how boundary information is propagated~\cite{tan2023scalable}. Across these variants, the important point is that windowing localizes the decoding problem while preserving enough buffer information to control boundary errors. The parallel regime then increases throughput by distributing window-level decoding tasks across classical workers.

These approaches do not remove feed-forward latency altogether. A window must still be processed before the corresponding committed information is available, and boundary handling introduces an additional response-time overhead. Their significance is that the overhead is bounded by the window size, buffer size, code distance, and available classical parallelism, rather than growing with the total experiment length. In memory experiments, windowing is often an efficiency strategy. In adaptive fault-tolerant computation, it becomes a mechanism for preventing the backlog growth described above.

Window-based decoding also clarifies an important distinction between local and global decoder outputs. Many graph-based decoders produce local correction information, local edge weights, or physical recovery operators that can be stitched across neighboring windows. A decoder that outputs only a single global logical bit for the entire experiment does not provide such an object. This distinction becomes especially important for neural decoders, because the most natural supervised learning target in a memory experiment is often a global logical label, whereas the most convenient object for window-based real-time decoding is local or window-local correction information.

A related hybrid strategy is to use a learned model as a pre-decoder rather than as a complete decoder. Chamberland~\emph{et al.}~\cite{chamberland2026fast} introduce an AI-based pre-decoder for the rotated surface code using a fully convolutional three-dimensional neural network. The model performs local spacelike and timelike corrections on the syndrome volume and passes the residual syndrome to a downstream global decoder such as MWPM or Union-Find. This differs from conventional neural decoders that consume the full syndrome history and output only the final logical class. The purpose of the pre-decoder is not to replace the algorithmic decoder, but to reduce the syndrome density of the instance that the algorithmic decoder must process. Since the runtime of MWPM and union-find depends strongly on syndrome density, this can reduce total decoding time while preserving the modularity of standard decoding pipelines. Because the pre-decoder is local and outputs residual syndromes rather than a global logical prediction, it also fits naturally with sliding-window and parallel-window decoding schemes.
\subsection{Neural architectures for real-time decoding}
\label{subsec:5.3}

The real-time behavior of a neural decoder is strongly tied to the architectural choices discussed in Sec.~\ref{sec:3}. Recurrent networks process syndrome data in a streaming manner and therefore fit naturally with online decoding, but their hidden state must be updated sequentially in time, which limits temporal parallelism. Graph neural networks exploit the locality of the detector graph or Tanner graph and are naturally compatible with local correction information, although long-range correlations require multiple message-passing rounds. Transformers provide direct access to long-range correlations through self-attention, but full attention over a spacetime syndrome volume has quadratic cost in the number of tokens. State-space architectures such as Mamba offer a different trade-off: they retain sequence memory through a scan-like state update and can reduce the dominant sequence-processing cost relative to attention-based models~\cite{gu2021efficiently,gu2023mamba}.

Conventional discriminative neural decoders were primarily designed for memory-experiment benchmarks. In this regime, it is natural to train a model to consume a complete syndrome history and output a single logical prediction, such as whether the final logical observable has flipped. This global-output formulation is powerful because it allows end-to-end learning from simulated or experimental data, including hardware-specific correlations, leakage, crosstalk, and other imperfections that may be difficult to model explicitly. AlphaQubit is a representative example: it uses a recurrent Transformer-based architecture to predict the logical correction for a complete memory experiment and can be fine-tuned on hardware data~\cite{bausch2024learning}. However, such global-output structure is not naturally suited to parallel-window decoding. A model that outputs only the final logical error does not produce a local physical correction, per-qubit marginal, boundary variable, or window-local contribution that can be merged across window seams. 

Recent neural-decoder work has therefore followed two complementary directions. The first is streaming acceleration: the neural architecture is redesigned so that inference can keep pace with the incoming syndrome stream. AlphaQubit~2 follows this route by using a scalable spatiotemporal architecture with temporal compression and accelerator-friendly spatial mixing~\cite{senior2025scalablerealtimeneuraldecoder}. Its real-time variant, AQ2-RT, reports sub-microsecond average throughput for intermediate-distance topological codes on commercial accelerators under the benchmarked conditions~\cite{senior2025scalablerealtimeneuraldecoder}. This should be interpreted as evidence that compact neural architectures can approach the throughput requirements of fast QEC cycles. It does not by itself imply that every end-to-end feed-forward latency constraint of a deployed control system is automatically satisfied.

Mamba-based neural decoders pursue a related acceleration strategy by replacing pairwise attention with a state-space sequence model~\cite{lee2025scalable}. For a distance-$d$ surface-code decoder that processes $m=O(d^2)$ spatial tokens per round, full attention over a single round has $O(m^2)=O(d^4)$ pairwise mixing cost, up to model-width factors. A selective state-space update can instead scale approximately linearly in the number of tokens, again up to model-width factors. The practical value of this change depends on implementation details, memory movement, and accelerator utilization, but the architectural motivation is clear: reduce the dominant sequence-processing cost while retaining long-range context.

Another strategy is to use strongly local, geometry-aware convolutions. Gu~\emph{et al.}~\cite{gu2026scalable} introduced  a convolutional neural decoder that exploits locality, translation equivariance, and direction-dependent message passing. For surface codes, the model uses three-dimensional convolutions over the spacetime syndrome lattice. The network embeds detection events, applies a stack of bottleneck convolutional blocks with depth scaled with the code distance, scatters the learned representations to data qubits, pools over the support of each logical observable, and outputs logical-error logits. This architecture remains a global logical-observable decoder, so it should not be viewed as directly solving the composability problem. Its relevance to real-time decoding is instead hardware regularity: the computation is local, feed-forward, fixed-depth, and amenable to low-precision implementation. The reported GPU benchmarks suggest favorable amortized throughput for buffered or batched decoding, while the authors also emphasize that meeting microsecond-scale superconducting budgets would require further optimization, such as depthwise convolutions, lower-precision arithmetic, or dedicated FPGA/ASIC implementations~\cite{gu2026scalable}.

The second direction is to make neural decoding compatible with parallel-window decoding by changing the supervision target. Instead of training one neural network to output a single logical bit for the entire experiment, one trains a model to output the logical contribution of a local temporal window. If the full syndrome history is decomposed into $M$ windows and the $j$th window outputs a bit $\hat{y}_j$, the final logical correction is obtained by
$
    \hat{y}
    =
    \hat{y}_1 \oplus \hat{y}_2 \oplus \cdots \oplus \hat{y}_M .
$
Zhang~\emph{et al.}~\cite{zhang2026learning} implements this idea by deriving per-window labels from a consistent set of local corrections and by jointly training across different window types. The subtlety is degeneracy near window boundaries. Two locally different correction chains may be globally equivalent, so there may be no unique canonical way to assign the logical contribution to one side of a seam. The neural-window approach addresses this by encouraging neighboring windows to adopt mutually consistent conventions during training. In this sense, the decoder learns a convention for decomposing the global logical correction into window-local contributions.

These two directions solve different problems. Streaming acceleration attempts to make a neural decoder fast enough to process the incoming syndrome stream directly. Parallel neural-window decoding changes the output structure so that many smaller neural decoding tasks can be distributed across classical workers. A scalable neural decoder may ultimately combine both ideas: a compact streaming architecture used inside a windowed or hierarchical decoding framework.

\subsection{Hardware-dependent timing budgets and deployment}
\label{subsec:5.4}

The practical timing budget is strongly platform dependent. Superconducting surface-code processors impose especially stringent constraints because QEC cycles can be on the order of microseconds. In Google's superconducting surface-code memory experiment, each QEC cycle lasted approximately $1.1\,\mu\mathrm{s}$, while the reported average decoder latency at distance five was $63\,\mu\mathrm{s}$~\cite{acharya2024quantum}. These numbers illustrate that online decoding and one-cycle feed-forward latency are distinct requirements.

Other hardware platforms provide different timing windows. In a trapped-ion Steane-code demonstration of real-time fault-tolerant error correction, each QEC cycle was reported to take less than $200\,\mathrm{ms}$~\cite{ryan2021realization}. In recent neutral-atom surface-code experiments, each QEC round was $4.45\,\mathrm{ms}$, while deeper logical teleportation layers were reported at $41.9\,\mathrm{ms}$~\cite{bluvstein2026fault}. These numbers should be read as representative experimental operating points rather than universal architectural constants. They show that the available classical-processing window can vary by several orders of magnitude across platforms.

This variation changes the decoder-design problem. Superconducting architectures place strong pressure on ultra-fast inference, low-latency data movement, and tight integration with cryogenic or room-temperature control electronics. Trapped-ion and neutral-atom systems may provide more time per QEC round, but they introduce different bottlenecks, including optical readout, image processing, atom motion, rearrangement, and mid-circuit feedback. Even in slower platforms, scalable decoding remains important once one accounts for large code distances, many logical patches, and repeated adaptive operations.

The choice of classical hardware is therefore part of the decoder architecture. GPUs and TPUs provide high throughput and are attractive for large neural models, but data movement and batching can be limiting when low-latency single-shot decisions are required. FPGAs and ASICs can provide more deterministic timing and tighter integration with control electronics, but they favor compact models, fixed-point arithmetic, limited memory access, and regular computation patterns. These constraints interact directly with the neural architecture. RNNs require sequential state updates, GNNs require message-passing schedules over sparse graphs, Transformers require attention kernels and substantial memory bandwidth, and state-space models require efficient scan or recurrence implementations.

For ML-based decoding, the relevant benchmark is therefore not simply whether a neural decoder achieves a lower logical error rate than MWPM, union-find, or a tensor-network decoder on an offline dataset. The relevant object is a hardware-dependent trade-off among logical error rate, sustained throughput, reaction latency, memory footprint, code-distance scaling, and robustness to hardware drift. A high-accuracy neural decoder that requires offline post-processing may be valuable for characterization and calibration, but it is not by itself a real-time decoder. Conversely, a compact streaming model may be useful even if it sacrifices some accuracy, provided it enables adaptive logical operations within the available timing budget.

The broader lesson is that real-time decoding is a quantum-classical co-design problem. The code, syndrome-extraction circuit, logical-operation schedule, decoder output representation, neural architecture, accelerator, and control protocol must be chosen together. This systems-level perspective is especially important for learned decoders, whose practical value is determined not only by expressive power and logical error suppression, but also by whether their inference pattern can be integrated into the timing constraints of a fault-tolerant quantum computer.

\section{Open Challenges and Future Directions}
\label{sec:6}
Recent progress has substantially strengthened the case for machine-learning-based quantum error decoding. Neural decoders have begun to achieve competitive or near-optimal accuracy on realistic surface-code and color-code benchmarks, and real-time variants have demonstrated that learned models can, in some regimes, keep pace with the measurement cycle of superconducting hardware. Nevertheless, these results should be viewed as evidence of a promising path rather than as a complete solution. Practical fault tolerance requires a decoder to satisfy several demanding requirements simultaneously: extremely low logical error rates, scalability to long computations, robustness to hardware-specific noise, and sufficiently low throughput and latency for real-time feedback.

A central challenge is the accuracy--latency trade-off. Larger and more expressive models tend to improve logical error suppression, but they also increase inference cost. Conversely, compact real-time models can meet stringent throughput targets only over a limited range of code distances, often with some loss in accuracy. This trade-off is especially important because the relevant target is not merely outperforming a baseline decoder, but reaching logical error rates required by large-scale fault-tolerant algorithms. Extending real-time neural decoding to the larger and more demanding operating regimes needed for practical computation remains a major open milestone. Future progress will likely require architecture innovation, model compression, low-precision inference, accelerator-specific optimization, and eventually deployment on specialized hardware such as FPGAs or ASICs.

Data efficiency, training complexity, and reliable evaluation are equally important. At low physical error rates, the training distribution is highly imbalanced: most samples contain no detection events or only simple low-weight error patterns, while ambiguous configurations near the decoding decision boundary are rare. Below threshold and at large code distance, logical failures themselves become rare events, making direct estimation of very small logical error rates prohibitively expensive with naive Monte Carlo sampling. Although trained neural decoders may have efficient inference once deployed, their training scalability remains a fundamental open question. Optimal decoding is computationally intractable in general, and a learned decoder should not be expected to evade this difficulty uniformly across arbitrary codes, noise models, and code distances. Even in structured physical settings, increasing the code distance requires the model to resolve more subtle distinctions between different logical error classes, often from increasingly rare informative samples. This suggests that training near-optimal ML-based decoders at progressively larger distances may require rapidly growing datasets, strong inductive biases, targeted data-generation strategies, or hybrid integration with algorithmic decoders. A pragmatic view is therefore that ML-based decoders may be most useful as highly optimized decoders for fixed, hardware-relevant code distances. Once such a distance is chosen, further reductions in logical error rates may rely primarily on lowering the physical error rate, improving device uniformity, exploiting hardware-specific noise structure, and co-designing the decoder with the physical platform.

Generalization remains another fundamental issue. A practical decoder should not merely interpolate within the distribution on which it was trained, but should remain reliable under realistic variations in factors such as nearby code distances, experiment duration, noise strength, calibration conditions, device drift, and logical operations. Recent neural decoders show encouraging signs of generalization across time and noise levels, but a systematic understanding of when such generalization occurs, how it depends on architecture and training distribution, and how it fails under hardware drift or distribution shift remains limited.

Another important direction is to move beyond standard memory experiments. Most neural decoders are first developed and benchmarked on quantum memory. Fault-tolerant computation, however, requires decoding during logical operations. Important examples include lattice surgery, code deformation, teleportation-based logical gates, and magic-state distillation. These settings introduce changing code geometries, time-dependent stabilizer structure, correlated faults across logical qubits, postselection or acceptance decisions, and classical feedforward constraints. Magic-state distillation, for example, naturally leads to correlated decoding across multiple logical blocks, while lattice-surgery-based computation requires decoding over dynamically changing space--time layouts rather than a fixed surface-code patch. Extending ML-based decoders from memory benchmarks to full computational workflows is therefore a crucial next step.

Real-time deployment also involves more than average throughput. A decoder may process syndrome data at the required average rate while still leaving a non-negligible delay between the final measurement and the final decoding result. This final-decision latency is particularly important for adaptive fault-tolerant protocols, where subsequent operations depend on decoded measurement outcomes. Moreover, the decoder must interface with measurement electronics, classical communication, scheduling, compilation, and Pauli-frame tracking. Thus, decoding should not be designed in isolation from the broader control stack. The decoder, compiler, scheduler, and classical control hardware together form a coupled real-time system.

% The scope of ML-based decoding will also need to broaden beyond the surface-code-centered setting that has dominated early benchmarks. Surface codes and related topological codes provide a natural and hardware-relevant testbed, but quantum low-density parity-check (qLDPC) codes are increasingly important because of their potential for lower overhead and finite-rate encoding. Their decoding problems differ substantially from surface-code decoding: the Tanner graph may be nonlocal, the number of logical qubits can be large, and direct classification over logical cosets can become infeasible. Developing ML methods that exploit the graph structure of qLDPC codes and scale to many logical qubits is an important future direction.

The scope of ML-based decoding will also need to broaden beyond the surface-code-centered setting that has dominated early benchmarks. Surface codes and related topological codes provide a natural and hardware-relevant testbed, but qLDPC codes are increasingly important because they offer a potential route to lower-overhead fault tolerance through finite-rate code families. Recent work on geometry-aware neural decoding for bivariate bicycle codes suggests that ML architectures can exploit regularity beyond planar surface-code geometry~\cite{gu2026scalable}, but it also highlights the need for code-specific inductive biases, scalable training procedures, and careful accuracy--latency benchmarking. Developing ML methods that exploit the graph or geometric structure of qLDPC codes, scale to many logical qubits, and interface effectively other algorithmic decoders is an important future direction.

Finally, ML-based decoding should be viewed as complementary to, rather than a replacement for, existing decoding methods. Hybrid approaches that combine learned components with graph-based, matching-based, belief-propagation, ordered-statistics, tensor-network, or search-based methods may offer favorable accuracy--latency trade-offs and better interpretability. Neural pre-decoders provide one example of this direction~\cite{chamberland2026fast}: a learned local model can reduce the density and complexity of the syndrome data before a conventional global decoder performs the final correction. More broadly, the most promising future direction may be a systems-level co-design perspective: designing quantum codes, logical-operation protocols, decoding algorithms, neural architectures, training distributions, compilers, and classical hardware together. Such an approach will be essential for turning learned decoders from successful demonstrations into reliable components of scalable fault-tolerant quantum computers.

\section{Conclusion}
\label{sec:7}

Machine learning provides a natural and increasingly powerful framework for quantum error decoding. Because decoding is fundamentally a problem of inferring logical error information from structured classical syndrome data, ML methods can leverage spatial locality, temporal correlations, graph structure, and hardware-specific noise patterns in ways that are often difficult to capture with purely hand-designed rules. As this chapter has emphasized, the resulting landscape is broad: discriminative, generative, and reinforcement-learning formulations offer different views of the decoding task, while recurrent, graph-based, attention-based, and state-space architectures are naturally suited to different kinds of structure in the data and therefore differ in their expressivity, scalability, latency, and suitability for particular code families, noise models, and input representations.

At the same time, ML-based decoding should not be viewed as a solved problem or as a uniform replacement for traditional decoders. The central challenge is to combine high accuracy with practical deployability under the stringent requirements of fault-tolerant quantum computation. This includes not only performance on standard memory benchmarks, but also robustness to realistic hardware noise, compatibility with real-time control, and eventual extension to logical operations beyond memory experiments. More broadly, quantum error decoding should also be viewed as a challenging and potentially fruitful problem for the ML community itself, since it combines rare but consequential failure events, strong spatiotemporal structure, distribution shift between simulation and hardware, and strict latency constraints in a way that is unusual within mainstream ML benchmarks. One promising direction is therefore a systems-level perspective, in which learned decoders are developed together with code structure, algorithmic decoding methods, hardware accelerators, and control architectures. In this sense, ML is not merely an additional decoding tool, but part of a broader co-design effort toward scalable fault-tolerant quantum computing.

\section*{Acknowledgements}
This work is supported by Institute of Information \& communications Technology Planning \& evaluation (IITP) grant funded by the Korea government (No. 2019-0-00003, Research and Development of Core Technologies for Programming, Running, Implementing and Validating of Fault-Tolerant Quantum Computing System), and the National Research Foundation of Korea (RS-2025-02309510).

\section*{Competing Interests}
The authors have no conflicts of interest to declare that are relevant to the content of this chapter.

%\bibliography{reference}
%\bibliographystyle{unsrt}

\end{document}